\documentclass[journal,twocolumn]{IEEEtran}
\ifCLASSINFOpdf
\else
\fi
\usepackage{amsfonts,amsmath,amssymb}
\usepackage{mathtools}
\usepackage{mdwmath}
\usepackage{cuted}
\usepackage{mdwtab}
\usepackage{amsthm}           
\usepackage{bm}               
\usepackage{units}            
\usepackage{graphicx}
\usepackage{epstopdf}
\usepackage[linesnumbered,ruled,vlined]{algorithm2e}
\usepackage{multicol}

\usepackage{url}

\usepackage{paralist}
\usepackage{color}
\usepackage{authblk}
\usepackage{cleveref}
\usepackage{float}
\usepackage[caption = false]{subfig}
\usepackage[nolist,nohyperlinks]{acronym} 

\newcommand{\Fig}[1]{Fig.~\ref{fig:#1}}

\newcommand{\Eq}[1]{Eq.~(\ref{eq:#1})}

\newcommand{\Lc}{\mathcal{L}}

\newcommand{\Rs}{\mathbb{R}^2}

\newcommand{\Ed}{\mathbb{E}}

\newcommand{\Pd}{\mathbb{P}}
\newcommand{\sinr}{\mathrm{SINR}}

\newcommand{\dr}{\mathrm{d}}

\newcommand{\Ru}{v(\gamma,r_s)}
\newcommand{\Rl}{u(\gamma,r_s)}


\begin{acronym}[NC-OFDM]
\acro{3gpp}[3GPP]{3\textsuperscript{rd} Generation Partnership Program}
\acro{5g}[5G]{Fifth Generation}
\acro{Adam}{Adaptive Moment Optimisation}
\acro{anr}[ANR]{Automatic Neighbour Relation}
\acro{ap}[AP]{Access Point}
\acro{atc}[ATC]{Air Traffic Control}
\acro{bs}[BS]{Base Station}
\acro{bpp}[BPP]{Binomial Point Process}
\acro{cc}[C\&C]{Command \& Control}
\acro{cdf}[CDF]{cumulative distribution function}
\acro{cdma}[CDMA]{Code Division Multiple Access}
\acro{cdr}[CD\&R]{Conflict Detection \& Resolution}
\acro{cfo}[CFO]{Carrier Frequency Offset}
\acro{comreg}[ComReg]{Commission for Communications Regulation}
\acro{cho}[CHO]{Conditional Handover}
\acro{dqn}[DQN]{Deep Q-Network}
\acro{ddqn}[DDQN]{Dueling Deep Q-Network}
\acro{csi}[CSI]{Channel State Information}
\acro{easa}[EASA]{European Aviation Safety Agency}
\acro{faa}[FAA]{Federal Aviation Administration}
\acro{fso}[FSO]{Free-Space Optical}
\acro{geo}[GEO]{Geosynchronous Equatorial Orbit}
\acro{gps}[GPS]{Global Positioning System}
\acro{gs}[GS]{Ground Station}
\acro{hap}[HAP]{High Altitude Platform}
\acro{iot}[IoT]{Internet of Things}
\acro{ipnn}[IPNN]{Interference Prediction Neural Network}
\acro{kpi}[KPI]{Key Performance Indicator}
\acro{lap}[LAP]{Low Altitude Platform}
\acro{leo}[LEO]{Low Earth Orbit}
\acro{los}[LoS]{Line-of-Sight}
\acro{lte}[LTE]{Long Term Evolution}
\acro{lstm}[LSTM]{Long Short-term Memory}
\acro{mac}[MAC]{Media Access Control}
\acro{mcp}[MCP]{Matern Cluster Process}
\acro{mc}[MC]{Monte Carlo}
\acro{3gpp}[3GPP]{3rd generation partnership project}
\acro{mimo}[MIMO]{Multiple Input Multiple Output}
\acro{mip}[MIP]{Mixed-Integer Programming}
\acro{mm}[MM]{Mapping Mechanism}
\acro{ml}[ML]{machine learning}
\acro{mno}[MNO]{Mobile Network Operator}
\acro{milp}[MILP]{Mixed-Integer Linear Programming}
\acro{nlos}[NLoS]{non-Line-of-Sight}
\acro{nn}[NN]{Neural Network}
\acro{nrt}[NRT]{Neighbour Relation Table}
\acro{ofdma}[OFDMA]{Orthogonal Frequency Division Multiple Access}
\acro{oot}[OOT]{Out-of-Tree}
\acro{osi}[OSI]{Open Systems Interconnection}
\acro{otdoa}[OTDoA]{Observed Time Difference of Arrival}
\acro{ott}[OTT]{Over-The-Top}
\acro{pv}[PV]{photo-voltaic}
\acro{pdf}[pdf]{probability density function}
\acro{ppp}[PPP]{Poisson Point Process}
\acro{qos}[QoS]{Quality of Service}
\acro{rc}[RC]{Remote Control}
\acro{rl}[RL]{Reinforcement Learning}
\acro{reqiba}[REQIBA]{REgression and deep Q-learning for Intelligent UAV cellular user to Base station Association}
\acro{rss}[RSS]{Received Signal Strength}
\acro{se}[SE]{Spectral Efficiency}
\acro{sir}[SIR]{Signal-to-Interference Ratio}
\acro{sinr}[SINR]{Signal-to-Interference-and-Noise Ratio}
\acro{snr}[SNR]{Signal-to-Noise Ratio}
\acro{uav}[UAV]{Unmanned Aerial Vehicle}
\acro{ue}[UE]{User Equipment}
\acro{ula}[ULA]{Uniform Linear Array}
\acro{wsn}[WSN]{Wireless Sensor Network}
\acro{wsn}[WSN]{Wireless Sensor Network}
\acro{rv}[RV]{random variable}
\acro{ppp}[PPP]{Poisson point process}
\acro{pgfl}[PGFL]{point generation functional}
\acro{pdf}[PDF]{probability density function}
\end{acronym}

\begin{document}
%
\title{REQIBA: Regression and Deep Q-Learning for Intelligent UAV Cellular User to Base Station Association
\thanks{Copyright (c) 2015 IEEE. Personal use of this material is permitted. However, permission to use this material for any other purposes must be obtained from the IEEE by sending a request to pubs-permissions@ieee.org. }}
%
%
%

\author{Boris Galkin*,
        Erika Fonseca*,
        Ramy Amer*,
        Luiz~A. DaSilva*$^\dagger$, 
        and Ivana Dusparic*
}

\affil{* CONNECT- Trinity College Dublin, Ireland \\
       $\dagger$ Commonwealth Cyber Initiative, Virginia Tech, USA \\
\textit{E-mail: \{galkinb,fonsecae,ramyr,duspari\}@tcd.ie, ldasilva@vt.edu}}

\maketitle

\begin{abstract}
\acp{uav} are emerging as important users of next-generation cellular networks. By operating in the sky, \ac{uav} users experience very different radio conditions than terrestrial users, due to factors such as strong \ac{los} channels (and interference) and \ac{bs} antenna misalignment. As a consequence, the \acp{uav} may experience significant degradation to their received quality of service, particularly when they are moving and are subject to frequent handovers. The solution is to allow the \ac{uav} to be aware of its surrounding environment, and intelligently connect into the cellular network taking advantage of this awareness. In this paper we present \ac{reqiba}, a solution that allows a \ac{uav} flying over an urban area to intelligently connect to underlying \acp{bs}, using information about the received signal powers, the  \ac{bs} locations, and the surrounding building topology. We demonstrate how \ac{reqiba} can as much as double the total \ac{uav} throughput, when compared to heuristic association schemes similar to those commonly used by terrestrial users. We also evaluate how environmental factors such as \ac{uav} height, building density, and throughput loss due to handovers impact the performance of our solution.
\end{abstract}


\vspace{-3mm}
\section{Introduction}
\label{sec:Introduction}

\acresetall

\acp{uav} are aircraft that operate without a pilot on board. Instead, they are either piloted remotely by a human operator, or they are controlled by computer algorithms. These devices are becoming increasingly used in a variety of applications, such as medical deliveries \cite{transplant}, building inspections, and surveillance \cite{8660516}. The Covid-19 pandemic in particular has accelerated the worldwide adoption of \acp{uav} for a variety of important use-cases, from disinfecting areas \cite{Gupta_2021} to enforcing social distancing in crowds \cite{Shao_2021}. Outside of those emergency use-cases, businesses are beginning to use \acp{uav} to deliver everyday items such as groceries \cite{manna_aero}, as well as for infrastructure inspections \cite{inspections}.
To enable these applications, the \acp{uav} will require a reliable wireless data link with their pilot or other controlling entities. While current commercially-available \acp{uav} rely on visual \ac{los} connections to their pilot, there is growing interest in connecting the \acp{uav} via  cellular networks \cite{3GPP_2018,amer2019mobility,8756296}. The emerging \ac{5g} family of cellular standards is intended to accommodate new types of users which require very high levels of reliability; this makes the \ac{5g} network an attractive option for facilitating \ac{uav} connectivity \cite{Ge_2019}. 

Until very recently, the cellular network was exclusively used by devices operating at -- or close to -- ground level. Existing cellular networks were designed with these users in mind, with \ac{bs} locations chosen to create coverage "cells" for the ground users, and the antennas configured to transmit signals towards the ground. Because they operate in the sky, \acp{uav} experience very different radio conditions to those of ground users, and the design of existing cellular networks introduces significant issues for them \cite{3GPP_2018}. Experimental trials have shown that while flying, \acp{uav} are likely to receive sidelobe signals from the \acp{bs}, as the mainlobes are aimed towards the ground \cite{Lin_2017}. The sidelobe signal gain may be such that a \ac{uav} may receive a stronger signal from a \ac{bs} which is kilometers away from it than from a \ac{bs} which is closer. Furthermore, a \ac{uav} is able to establish unobstructed \ac{los} channels to a large number of \acp{bs}. The consequence of this is that, while at ground level the network may be partitioned into coverage "cells", at greater heights the network coverage is highly volatile, with very strong interference from distant \acp{bs} and a large number of \acp{bs} that a \ac{uav} can connect to at any given moment \cite{3GPP_2017}. In our prior experimental work \cite{galkin_2021}, we demonstrated that a \ac{uav} flying in a city can successfully connect via strong side-lobe signals to small cells up to a kilometer away.

As it travels through an area, a \ac{uav} is likely to see very rapid signal fluctuations, and may potentially hand off from one \ac{bs} to another very frequently. These frequent handovers may introduce significant overheads into the network performance, and degrade the service quality for the \ac{uav} link. In our prior work \cite{mobility-challenges} we discussed some of the solutions that could be implemented by network operators to mitigate these handover issues, such as adjusting the \ac{anr} mechanisms used by neighbouring \acp{bs} to take into account the unique flight behaviours of \acp{uav}. In addition to changes made on the side of the network by the operators, the \acp{uav} themselves can be designed to tackle some of these connectivity issues.
The \ac{3gpp} has suggested that steerable, directional antennas should be used by the \acp{uav}, as they can allow a \ac{uav} to improve its wireless link quality by reducing the power of undesirable \ac{bs} signals (i.e. interference) \cite{3GPP_2018}. A number of works have suggested that \acp{uav} should optimise their flight trajectories with respect to the underlying cellular network, to improve performance. A variety of algorithms have been proposed for this trajectory optimisation, as discussed in the next section. While this type of optimisation is useful for scenarios where the \ac{uav} trajectory can be optimised with respect to cellular service, there are a variety of scenarios where the trajectory of the \ac{uav} may not be modified, either because the flight path is explicitly defined by the \ac{uav} mission (such as photography work) or because the \ac{uav} is being piloted in real-time by a human operator rather than a computer algorithm. Furthermore, algorithm-driven \ac{uav} swarm flights are currently heavily restricted under European Union (EU) regulations, and it is not clear when these restrictions will be lifted \cite{EASA_2019}. As such, we expect that human-piloted \acp{uav} will make up a large part of the \ac{uav} landscape in the near future. 

For these reasons, there is a need to explore intelligent \ac{uav} cellular connectivity that does not rely on optimising \ac{uav} trajectories. In these scenarios, the \ac{uav} can improve its service quality and manage its handover rate by intelligently choosing which \acp{bs} to associate with, using knowledge of its surrounding environment. When equipped with a steerable directional antenna, this would allow the \ac{uav} to align its antenna to create the best wireless channel for the given circumstances. This association decision will be complicated by the volatile radio environment discussed above, with factors such as the \ac{uav} mobility, the large number of candidate \acp{bs} with strong side-lobe signals, and the impact of \ac{los}-blocking buildings all affecting the decision of which \ac{bs} to connect to, and when to carry out a handover.

In our prior work \cite{Galkin_2020} we addressed the issue of intelligent \ac{uav}-\ac{bs} association in a static scenario where the \ac{uav} was hovering in place and needed to make a single association decision for its location. While our proposed neural network solution was shown to outperform conventional association schemes in terms of channel quality, as the scenario was static we did not address the issue of \ac{uav} movement and the resulting handovers. In this paper we extend our prior work by considering a scenario where a \ac{uav} needs to intelligently maintain a connection to the underlying cellular network while moving, by making multiple association decisions during flight. Our contributions are as follows:

\begin{itemize}
    \item We propose a novel neural network-based solution which we refer to as \ac{reqiba}, which allows a \ac{uav} equipped with a directional antenna to intelligently associate with nearby \acp{bs} during flight. This solution consists of a regression neural network and a \ac{ddqn} module. The solution takes in state information about the environment, information about received signal power, interference, and current \ac{bs} connection. The network then chooses a \ac{bs} to connect to based on these factors, to maximise the data throughput.
    \item We evaluate the performance of our \ac{reqiba} solution and demonstrate that addressing this problem in a mobility scenario is indeed a lot more complex than treating it as a series of static connection decisions. We show how \ac{reqiba} outperforms our prior solution in \cite{Galkin_2020} by simultaneously increasing the total throughput and reducing the rate of \ac{uav} handovers. 
    \item We compare the performance of \ac{reqiba} against heuristic association schemes such as those found in the literature. We demonstrate how it outperforms these heuristic schemes under different environmental conditions, while exploring how these environmental conditions affect its performance improvement.
\end{itemize}

This paper is structured as follows. In Section \ref{sec:Related} we review the related works. In Section \ref{sec:SystemModel} we outline our system model. In Section \ref{sec:ML} we introduce and describe our \ac{reqiba} solution. In Section \ref{sec:Methodology} we describe how \ac{reqiba} is trained and evaluated. In Section \ref{sec:Results1} we compare the performance of \ac{reqiba} against our prior solution in \cite{Galkin_2020}. In Section \ref{sec:Results2} we evaluate how \ac{reqiba} performs against the heuristic algorithms under various environmental conditions. Finally, in Section \ref{sec:Conclusion} we provide our conclusions and discuss directions for further investigation.

\vspace{-3mm}
\section{Related Works}
\label{sec:Related}
The cellular connectivity issues experienced by flying \acp{uav} 
have been extensively explored. In \cite{qualcomm-sim} Qualcomm reports the results of a series of simulations and field measurements which determine that \acp{uav} are exposed to stronger interference than ground users. In \cite{Azari_2019} the authors apply stochastic geometry to demonstrate how \acp{uav} experience  throughput degradation with increasing heights, due to growing interference power. In \cite{nokia-sim,Euler_2019} the authors use simulations to show how \acp{uav} experience very high handover failure rates due to strong interference conditions at large heights. In \cite{HCC:3325421.3329770}, authors perform an experimental flight with a \ac{uav} at different heights and speeds, and conclude that a \ac{uav} performs approximately 5 times more handovers than ground users moving at the same speed. In our prior work \cite{Galkin_2018} we apply stochastic geometry to analytically characterise the coverage probability and throughput of a \ac{uav} that is connected to a network of terrestrial \acp{bs} via  directional antenna. In our prior work \cite{amer2020performance} we demonstrate how \ac{bs} sidelobes can cause frequent \ac{uav} handovers during \ac{uav} vertical movement. 

As \acp{uav} are highly mobile devices, the wireless community typically approaches the problem of \ac{uav} cellular connectivity from the perspective of optimising the \ac{uav} trajectory. In \cite{trajec-mobile} the authors consider a 
\ac{uav} that needs to fly between two locations, in a manner that minimises the flight time while maintaining a reliable cellular link. They use a graph representation of the network and apply Dijkstra's algorithm to find the route of the \ac{uav}. The authors of \cite{Chen_2017} optimise the movement of a \ac{uav} around a map of \ac{los}-blocking buildings, to ensure the \ac{uav} maintains a \ac{los} channel to its \ac{bs}. A similar work is carried out in \cite{Esrafilian_2018}, where \ac{uav} relays are intelligently positioned around known user locations as well as the locations of buildings. In \cite{Gangula_2018} the authors optimise the \ac{uav} trajectory given available landing sites where the \ac{uav} can land and reduce its energy consumption. As there is significant research interest in the \ac{uav} trajectory optimisation topic, \ac{ml} is seeing widespread application in the \ac{uav} domain. In our prior work \cite{omoniwa_2021} we develop a multi-agent ML scenario where multiple \ac{uav} \acp{bs} are deployed to provide service in a disaster scenario. Using distributed multi-agent Q-Learning the \acp{uav} optimise their placement above the user locations while simultaneously reducing their energy consumption. In \cite{Huang_2020} the authors consider a \ac{dqn} algorithm to optimise the trajectory of multiple \acp{uav} in a way that ensures that they have a sufficient connection to a terrestrial \ac{bs} with massive MIMO antennas. The authors of \cite{Liu_2020} consider a mobile edge computing scenario where a \ac{dqn} is applied to optimise a \ac{uav}'s trajectory and allow it to maximise the amount of computing tasks it offloads from ground users. In \cite{Xiong_2021} the authors consider an \ac{iot} scenario where a \ac{uav} relay applies a \ac{dqn} to adjust its trajectory around \ac{iot} devices, while wirelessly recharging its battery at certain locations. In \cite{challita2018deep}, authors investigate interference-aware trajectory optimisation using game-theory and \ac{ml} for the purpose of maximizing energy efficiency and minimizing wireless latency and the UAV interference on the ground network. Authors in \cite{path-rl} propose an \ac{ml} approach to find an optimal trajectory which minimizes the travel time while maintaining connectivity with the cellular network. Meanwhile, the work in \cite{zhong2020deep} proposes a deep learning-based framework to manage the dynamic movement of multiple \acp{uav} in response to ground user mobility so as to maximize the sum data rate of the ground users. It should be noted that the works described above assume interference-free channel conditions, omni-directional antenna radiation patterns of the \acp{uav} and \acp{bs}, guaranteed \ac{los} propagation conditions, or some combination thereof. This allows for tractable optimisation problems and simplifies the \ac{uav}-\ac{bs} association behaviour, but it may not reflect a realistic urban environment or \ac{uav} use-case.

Along with the ongoing work on \ac{uav} trajectory optimisation, the wireless community is also beginning to address the issues associated with \ac{uav} handovers using \ac{ml} tools. For instance, the authors in \cite{Chowdhury_2020} show how \ac{ml} can be used to dynamically adjust the \ac{bs} antenna tilt angles. The authors apply model-free \ac{rl} to the \ac{bs} antenna tilt such that the agent balances the received signal power for a \ac{uav} user passing overhead with the throughput of the ground users. The authors demonstrate how this intelligent antenna tilting can help reduce the \ac{uav} handover rate without significant performance loss for the ground users. In \cite{AAzari_2020}, the authors consider a joint \ac{bs} selection and resource allocation problem for moving \acp{uav}. The authors apply reinforcement learning to simultaneously select the  serving \ac{bs} and the allocated resource blocks with the aim of minimizing the uplink interference created by the \ac{uav} for the ground users, while keeping the rate of \ac{uav} handovers manageable. A similar problem is addressed in \cite{Chen_2020}, where the authors intelligently select \ac{bs} associations for a \ac{uav} moving along a known trajectory to minimise the rate of handovers. In \cite{takacs2020methods}, the authors envision a method for managing a \ac{uav} flight path to coordinate enhanced handover in \ac{3gpp} networks. In \cite{Du_2020} the authors propose using \ac{ml} to estimate channel conditions in noisy environments, as well as the use of \ac{lstm} networks to predict \ac{uav} trajectories in future timesteps for handover planning. In our prior work \cite{Galkin_2020}, we consider the problem of \ac{uav} association, where the \ac{uav} is equipped with a directional antenna for communication, and an omni-directional antenna for sensing. Our proposed \ac{ml} solution in that paper uses the available channel information from the omni-directional antenna as well as the known locations of the interfering \acp{bs} to infer which \ac{bs} will exhibit the best channel conditions for the directional antenna.

This paper extends our prior work in \cite{Galkin_2020}. While our prior work considers optimising the channel quality in a scenario where a \ac{uav} is hovering at a fixed location, in this work we consider a moving \ac{uav}. This movement introduces the issue of \ac{uav} handovers, which complicates the association problem and requires an entirely new \ac{ml} solution. Note that our work differs from existing works such as \cite{AAzari_2020} and \cite{Chen_2020} in that we consider a throughput-maximisation problem for a \ac{uav} which communicates via a steerable directional antenna rather than an omni-directional one. This complicates the process of gathering environmental information for the association decision, which requires us to use a more complex \ac{ml} solution to successfully optimise the \ac{uav} performance, as we will demonstrate in later sections. We further differentiate our work from the state-of-the-art by considering the effect of \ac{bs} antenna side-lobes, \ac{los} blockage due to buildings, as well as inter-cell interference from neighbouring \acp{bs}, all of which result in a more complicated radio environment. 

\section{System Model} 
\label{sec:SystemModel}

\begin{figure}[t!]
\centering
	\subfloat{\includegraphics[width=.45\textwidth]{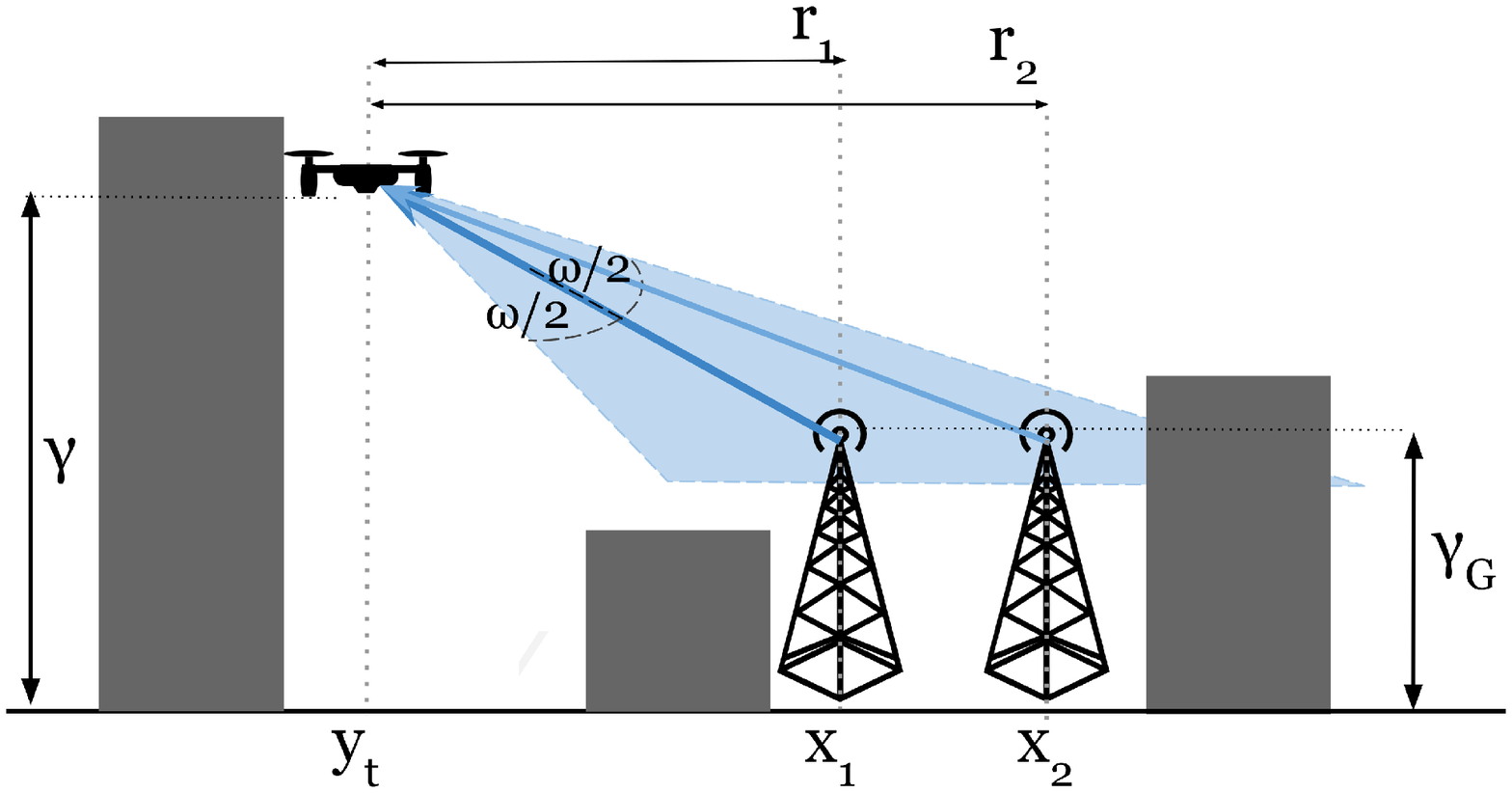}}\\
	\vspace{-7mm}
	\subfloat{\includegraphics[width=.45\textwidth]{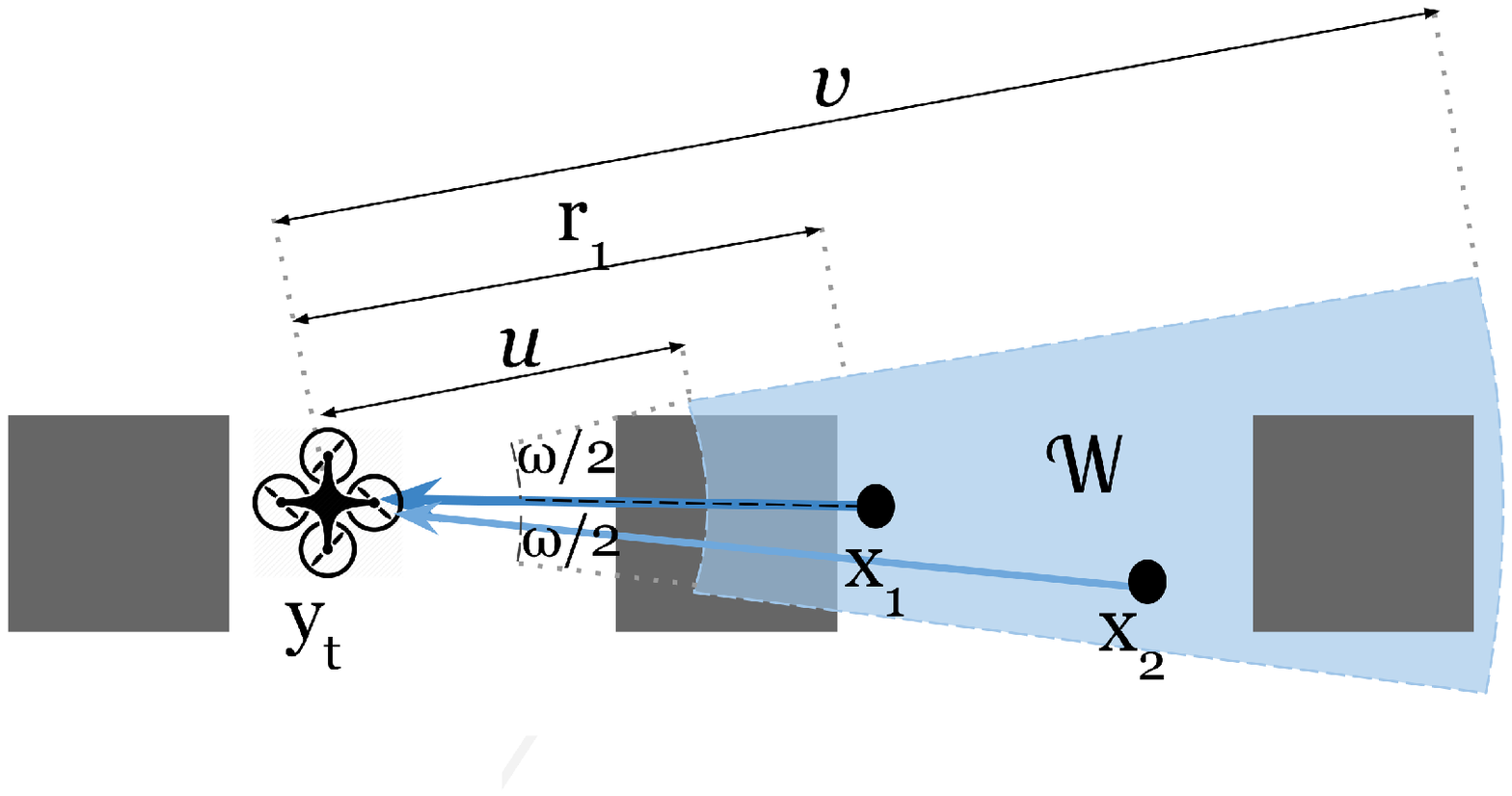}}
	\vspace{-9mm}
	\caption{
	Side and top view showing a UAV in an urban environment at a height $\gamma$, positioned above $y_t$ with antenna beamwidth $\omega$. The UAV chooses to associate with the \ac{bs} at $x_1$ and centers its antenna main lobe on the \ac{bs} location; the blue area $\mathcal{W}$ illuminated by the main lobe denotes the region where interferers may be found. The \ac{bs} at $x_2$ falls inside this area and produces interference. 
	\vspace{-5mm}
	}
	\label{fig:drone_network}
\end{figure}


We consider a scenario where a \ac{uav} is performing a mission in an urban environment, such as a package delivery. The \ac{uav} is being piloted remotely, either by a human operator or by a flight controller, with the wireless data connection being carried out via the underlying cellular network. We assume that the \ac{uav} flight trajectory is not planned with respect to the locations of the underlying cellular network infrastructure, either because the pilot entity is not aware of the infrastructure locations and performance, or because the \ac{uav} mission takes precedence when planning the \ac{uav} trajectory. The underlying cellular network consists of \acp{bs} which are horizontally distributed as a homogeneous \ac{ppp} $\Phi = \{x_1 , x_2 , ...\} \subset \Rs$ of intensity $\lambda$, at a height $\gamma_{G}$ above ground. Elements $x_i\in \Rs$ represent the projections of the \ac{bs} locations onto the $\Rs$ plane. The \ac{uav} travels from an initial location $y_1 \in \Rs$ to a final location $y_T \in \Rs$ in a straight line over a length of time T. We discretise time into $T$ timesteps: this lets us partition the travel vector of the \ac{uav} into $T$ coordinates at different timesteps in the journey. We define the 
vector of these \ac{uav} coordinates as 
$\mathbf{u}=(y_1,y_2,...y_t,...y_T)$,
where $y_t$ denotes the coordinates of the \ac{uav} in the $t$-th timestep. The \ac{uav} height above ground is assumed to remain constant throughout the flight due to factors such as \ac{uav} mission requirements or air traffic control guidelines. The \ac{uav} height is denoted as $\gamma$. For an exploration of intelligent cellular-\ac{uav} height adjustment the reader is referred to our prior work \cite{fonseca_2020}. Let $r_{i,t} = ||x_i-y_t||$ denote the horizontal distance between the coordinates $x_i$ and $y_t$, and let $\phi_{i,t} = \arctan(\Delta \gamma/r_{i,t})$ denote the vertical angle, where $\Delta \gamma = \gamma - \gamma_{G}$. 

The \ac{uav} is equipped with two sets of antennas: an omni-directional antenna for \ac{bs} pilot signal detection and signal strength measurement, as well as a directional antenna for communicating with the \ac{uav}'s associated \ac{bs}. The omni-directional antenna has an omni-directional radiation pattern with an antenna gain of 1, while the directional antenna has a horizontal and vertical beamwidth $\omega$ and a rectangular radiation pattern; using the antenna radiation model in \cite{8422376}, the antenna gain is given as $\eta(\omega) = 16\pi/(\omega^2)$ inside of the main lobe and $\eta(\omega)=0$ outside. We denote the coordinates of the \ac{bs} which the \ac{uav} is associated with at time $t$ as  $x_{s,t} \in \Phi$ and its horizontal distance to the \ac{uav} as $r_{s,t}$. Whenever the \ac{uav} connects to a \ac{bs} $x_{s,t}$ it aligns its directional antenna towards $x_{s,t}$; this results in the formation of an antenna radiation pattern around $x_{s,t}$ which we denote as $\mathcal{W} \subset \Rs$, as depicted in Fig. \ref{fig:drone_network}. This area takes the shape of a ring sector of arc angle equal to $\omega$ and major and minor radii $\Ru$ and $\Rl$, respectively, where

\begin{align}
\Ru = 
\begin{cases}
\frac{|\Delta \gamma|}{\tan(|\phi_{s,t}|-\omega/2)} \hspace{-2mm} &\text{if} \hspace{3mm} \omega/2 < |\phi_{s,t}| < \pi/2 - \omega/2 \\
\frac{|\Delta \gamma|}{\tan(\pi/2 -\omega)} \hspace{-2mm} &\text{if} \hspace{3mm} |\phi_{s,t}|  > \pi/2 - \omega/2 \\
\infty &\text{otherwise} \nonumber
\end{cases}
\end{align}

\begin{align}
\Rl = 
\begin{cases}
\frac{|\Delta \gamma|}{\tan(|\phi_{s,t}| +\omega/2)} \hspace{2mm} &\text{if} \hspace{3mm} |\phi_{s,t}|  < \pi/2 - \omega/2  \\ 
0 &\text{otherwise}
\end{cases}
\end{align}

\noindent
with $|.|$ denoting absolute value. The \acp{bs} which fall inside the area $\mathcal{W}$ are denoted by the set $\Phi_{\mathcal{W}} = \{x \in \Phi : x \in \mathcal{W}\}$. The \acp{bs} in the $\Phi_{\mathcal{W}}$ are capable of causing interference to the \ac{uav}-\ac{bs} communication link, as their signals may be received by the \ac{uav}'s directional antenna with non-zero gain.

As we are considering an urban environment, buildings will affect the wireless signals by blocking \ac{los} links. We model these buildings as being distributed in a square grid, following the model proposed by the ITU in \cite{ITUR_2012}. The density of buildings per square kilometer is $\beta$ and the fraction of the ground area covered by buildings 
is $\delta$. All buildings have the same horizontal dimensions, and each building has a height which is a Rayleigh-distributed random variable with scale parameter $\kappa$. The \ac{uav} will have an unobstructed \ac{los} channel towards a \ac{bs} $i$ at time $t$ if there exist no buildings which intersect the straight line between $x_i$ at height $\gamma_G$ and $y_t$ at height $\gamma$. Otherwise if there is at least one building that intersects this line then the channel is \ac{nlos}.

\noindent

We assume that the \acp{bs} have tri-sector antennas, with each antenna being a \ac{ula} with $N_t$ antenna elements. For tractability we model these antennas as being horizontally omni-directional with horizontal gain 1. The vertical gain of these antennas is a function of the angle between the \ac{uav} and the \ac{bs} and is defined similar to \cite{8756719} as

\begin{equation}
\label{antten-gain}
\mu(\phi_{i,t}) = \frac{1}{N_t}\frac{\sin^2 \frac{N_t \pi}{2}\sin(\phi_{i,t})}{\sin^2 \frac{\pi}{2}\sin(\phi_{i,t})} .
\end{equation}

When the \ac{uav} is connected to the \ac{bs} at $x_s$ at timestep $t$, the \ac{sinr} of the signal received by the directional antenna is given as

\begin{equation}
\sinr_t = \frac{p  \eta(\omega)\mu(\phi_{s,t}) c ((r_{s,t})^2+\Delta \gamma^2)^{-\alpha_{z_{s,t}}/2}}{I+\sigma^2} 
\label{eq:SINR}
\end{equation}

\noindent
 where $p$ is the \ac{bs} transmit power, $\alpha_{z_{s,t}}$ is the pathloss exponent, $z_{s,t} \in \{\text{L},\text{N}\}$ is an indicator variable which denotes whether the \ac{uav} has \ac{los} or \ac{nlos} to its serving \ac{bs} $x_s$ at timestep $t$, $c$ is the near-field pathloss, $\sigma^2$ is the noise power, and $I$ is the aggregate interference power received from the \acp{bs} in the set $\Phi_{\mathcal{W}}$. As we are considering discrete timesteps with durations in the order of seconds, we approximate the channels as having no fast-fading effects, since effects such as multipath fading are expected to average out over the given timescales \cite{Azari_2019,Chen_2017}. However, the instantaneous \ac{sinr} experienced by the \ac{uav} at a given moment may be affected by a number of factors, including multipath fading. If this multipath fading follows a Nakagami-m distribution the \ac{cdf} of the instantaneous \ac{sinr} can be derived mathematically. These derivations are given in the Appendix.
 
 The throughput per unit of bandwidth at timestep $t$ is given by the Shannon bound as 
 
 \begin{equation}
R_t = \log_2(1+\sinr_t).\label{eq:throughput}
 \end{equation}
 
 
 As already mentioned, we assume that the \ac{uav} points its directional antenna at the \ac{bs} it is currently associated with. The \ac{uav} is capable of seamlessly tracking the changing \ac{bs} orientation using its directional antenna as it moves. 
 If, however, the \ac{uav} changes its associated \ac{bs} in a timestep then the \ac{uav} will spend a portion of that timestep realigning its directional antenna towards the new \ac{bs}. We assume this antenna realignment, along with the handover signalling involved \cite{Azari_2019}, causes an overhead which reduces the effective throughput in that timestep. A number of factors will determine the extent of this overhead, such as the configuration of the network by the network operators, the handover protocols in place, and the design of the \ac{uav} itself. For example, the \ac{anr} mechanism in the \ac{bs} network may not be configured to support \ac{uav} users attempting handovers to distant \acp{bs}, in which case the handover process will be delayed with additional signalling \cite{mobility-challenges}. The network may adopt new, streamlined handover mechanisms to support 5G vehicular users \cite[Section VIII.B]{Garcia_2021}, which will also result in more streamlined handover signalling for the \ac{uav} user. Finally, handover delays may occur if the \ac{uav} directional antenna relies on alignment via the physical rotation of the \ac{uav} frame, which can take up to several seconds for a quadcopter style \ac{uav} \cite{matrice_specs}. To preserve the generality of our work, all of these factors are abstracted into the handover penalty $\tau \in [0,1]$, where smaller values of $\tau$ correspond to bigger performance impact; the exploration of the relationship between the aforementioned factors and the resulting performance overheads is left for a future work.
 
 Note that with tri-sector antenna \acp{bs} it is possible for a user to handover between two antennas of the same \ac{bs}; we assume that this type of handover occurs seamlessly and for this reason our focus in this paper is on the handovers between different \acp{bs}. 
 
 \vspace{-2mm}
\section{\ac{reqiba}}
\label{sec:ML}

  \subsection{Problem Statement}
Similar to the state-of-the-art, which considers \ac{uav} optimisation via decentralised algorithms, we assume that the \ac{uav} runs an algorithm on-board its hardware that allows it to make intelligent decisions based on observed information. The \ac{uav} is to fly through the environment from a starting point to an ending point over $T$ timesteps. At every timestep it is to make a decision about which \ac{bs} it should be connected to for that timestep. If it decides to connect to a \ac{bs} other than the one it is currently connected to, it will carry out a handover in that timestep. The reward function for the timestep $t$ is given as 
 
 \begin{equation}
 \hspace{-5mm}\rho_t = 
 \begin{cases}
 \log_2(1+\sinr_t) &\text{if no handover}  \\
 \tau \log_2(1+\sinr_t) &\text{otherwise}
 \end{cases}
 \label{eq:reward}
 \end{equation}

where $\tau$ is the handover penalty factor. It follows that the smaller the value of $\tau$ the less desirable it is for a handover to occur. 

The optimisation problem is stated as follows:
\begin{subequations}
\begin{align}
 &\underset{(x_{s,1},x_{s,2}...,x_{s,T})}{\text{max} } \sum_{t=1}^T \rho_t \\
 &\text{s.t.} \sum_{j=1}^{|\Phi|} \textbf{1}(x_{s,t}=x_j) = 1,\quad \forall t \in [1...T]
 \label{eq:constraint}
 \end{align}
 \end{subequations}

\noindent
where $|\Phi|$ denotes the cardinality of the set $\Phi$ and $\textbf{1}(.)$ denotes the indicator function, which equals $1$ when the expression within the parentheses is true. The constraint in \Eq{constraint} guarantees that the \ac{uav} must be connected to exactly one \ac{bs} in $\Phi$ at each timeslot in the episode.


In theory the \ac{uav} may choose from any \ac{bs} in $\Phi$ for the optimisation problem above, but from a practical perspective the choice tends to be limited to only a subset of those \acp{bs}. From our prior work \cite{Galkin_2020} we have observed that the \ac{uav} is likely to get the best connection from one of the closest \acp{bs} to it, or one of the \acp{bs} with the strongest received signal power. We therefore denote a subset of \acp{bs} $\Phi_\zeta \subset \Phi$, where $\Phi_\zeta$ consists of the $\zeta/2$ closest \acp{bs} to the \ac{uav} at timestep $t$, as well as the $\zeta/2$ \acp{bs} with the strongest received signal power at the \ac{uav} at that timestep (as measured by the omni-directional antenna). Therefore, at timestep $t$ the action the \ac{uav} takes is to choose from one of the $\zeta$ candidate \acp{bs} in $\Phi_\zeta$. Note that the $i$-th closest \ac{bs} at timestep $t$ may also be the $j$-th strongest signal \ac{bs} where $i,j \leq \zeta/2$, which means that 
the cardinality of set $\Phi_\zeta$ may be lower than $\zeta$. The omni-directional antenna on the \ac{uav} also allows it to select candidate \acp{bs} based on the highest omni-directional \ac{sinr}; however, since the received interference at the omni-directional antenna will greatly differ from the interference received by the directional antenna once aligned, the omni-directional \ac{sinr} of the candidate \acp{bs} is unlikely to be indicative of the directional antenna \ac{sinr} (this will be demonstrated in Section \ref{sec:Results2}). For this reason, we do not consider the omni-directional \ac{sinr} of the candidate \acp{bs} for the subset $\Phi_\zeta$.

 \subsection{Environment Information}

Before describing our proposed \ac{reqiba} solution we specify what state information is assumed to be available to the UAV for use in its decision-making:
 
 \begin{itemize}
     \item The \ac{bs} to which the \ac{uav} is currently associated. This is updated every time the \ac{uav} hands over to a new \ac{bs}.
     \item The received signal power from nearby \acp{bs}. The \ac{uav} has an omni-directional antenna for sensing the environment, being able to receive the pilot signals from nearby \acp{bs} and determine their received signal power. The received signal powers of all of the candidate \acp{bs} $\Phi_\zeta$ are measured in this manner.
     \item The 3D coordinates of the \ac{uav} and the \acp{bs} in $\Phi$. The \ac{uav} knows its location from its \ac{gps} coordinates, while the locations of the \acp{bs} are provided to it by the network. This location information is independent of the received signal power measurements carried out by the omni-directional antenna; as such, the \ac{uav} knows the locations of \acp{bs} even if they are too far away to have their pilot signals decoded by the antenna. This information is assumed to be error-free, and in the case of the moving \ac{uav} is updated in real-time.
     \item The 3D topology of the environment, as in \cite{Esrafilian_2018}. To safely navigate through the urban environment we assume the \ac{uav} has information on the locations and heights of buildings around it. This topology information is error-free, and assumed to be accurate enough to allow the \ac{uav} to determine whether it has an unobstructed \ac{los} channel to a given \ac{bs} using a ray trace.
     \item The directional and omni-directional antenna parameters. The \ac{uav} is aware of the gain patterns for both sets of its antennas, and is able to determine the area $\mathcal{W}$ that would be illuminated by the directional antenna should it point it towards a given location.
     \item Whether or not it will need to optimise its cellular link in the next timestep. We assume the \ac{uav} knows when it no longer needs to continue optimising its cellular connectivity (i.e., the end of the optimisation episode). This information is necessary to tell the learning algorithm whether it should consider future timesteps when choosing an action, or only the current timestep.
 \end{itemize}

As the \ac{uav} does not have access to important environmental information such as the channel propagation conditions or its trajectory in future timesteps, it is not possible to solve the \ac{bs} optimisation problem offline before the \ac{uav} flight. This motivates our choice of an online \ac{rl} approach.
\subsection{\ac{reqiba} Solution Structure}

It is clear from the reward function in \Eq{reward} and the \ac{sinr} expression in \Eq{SINR} that the  throughput $\rho_t$ at timestep $t$, which is obtained after the \ac{uav} chooses the \ac{bs} for that timestep, is affected by four factors: whether a handover occurs, the performance penalty for the handover, the received signal power from the associated \ac{bs}, and the interference power from the other \acp{bs}. 
The handover penalty and the interference powers are not known explicitly by the \ac{uav}, and thus cannot be used directly in the decision-making process. Recall that the omni-directional antenna that the \ac{uav} uses for sensing and the directional antenna used for communication have different radiation patterns, and that the \ac{uav} aligns its directional antenna towards its chosen \ac{bs}, with the interfering \acp{bs} coming from the area $\mathcal{W}$. The \ac{uav} omni-directional antenna is not able to measure the interference power that comes specifically from the area $\mathcal{W}$: this needs to be estimated in some other way. The \ac{uav} has access to the map of the \acp{bs}, and knows its own directional antenna beamwidth, so it can determine which \acp{bs} will fall inside the area $\mathcal{W}$ and cause interference.  

The inputs to our \ac{reqiba} solution are  as follows. The received signal powers of the $\zeta$ candidate \acp{bs} are provided as a vector $\mathbf{p}_\zeta = (p_{1,t},p_{2,t},...,p_{\zeta,t})$, where $p_{i,t} = p \mu(\phi_{i,t})c ((r_{i,t})^2+\Delta \gamma^2)^{-\alpha_{z_{i,t}}/2}$. The handover information is conveyed with a vector of binary flags $\mathbf{o}_\zeta = (o_{1,t},o_{2,t},...,o_{\zeta,t})$ where $o_{i,t} = 1$ if $x_{i,t} = x_{s,(t-1)}$ and 0 otherwise, to indicate which of the candidate \acp{bs} the \ac{uav} is currently associated with. Information about the interfering \acp{bs} consists of two $\zeta \times \xi$ input matrices $\mathbf{F_\zeta}$ and $\mathbf{L_\zeta}$, where $\xi$ denotes the number of interfering \acp{bs} to consider per link. $\mathbf{F_\zeta}$ contains the horizontal distances of the interfering \acp{bs} to the \ac{uav}, where the $i$-th row corresponds to the area $\mathcal{W}_i$ illuminated when the \ac{uav} points its directional antenna towards the $i$-th candidate \ac{bs}, and the $j$-th column represents the $j$-th closest interfering \ac{bs} in the corresponding illuminated area. The matrix $\mathbf{L_\zeta}$ contains binary flags to indicate whether the corresponding interfering \acp{bs} have \ac{los} or \ac{nlos} to the \ac{uav}, as determined from the building topology map \cite{Esrafilian_2018}. If the $i$-th candidate \ac{bs} has fewer than $\xi$ interferers then the remaining entries in the $i$-th rows of $\mathbf{F_\zeta}$ and $\mathbf{L_\zeta}$ take null values. The final two inputs are the \ac{uav} height above ground $\gamma$ and a binary flag $\mathbf{t}$ which takes a value of 1 if the current timestep is the final timestep in the episode.

\begin{figure*}[t!]
\centering
	\includegraphics[width=.70\textwidth]{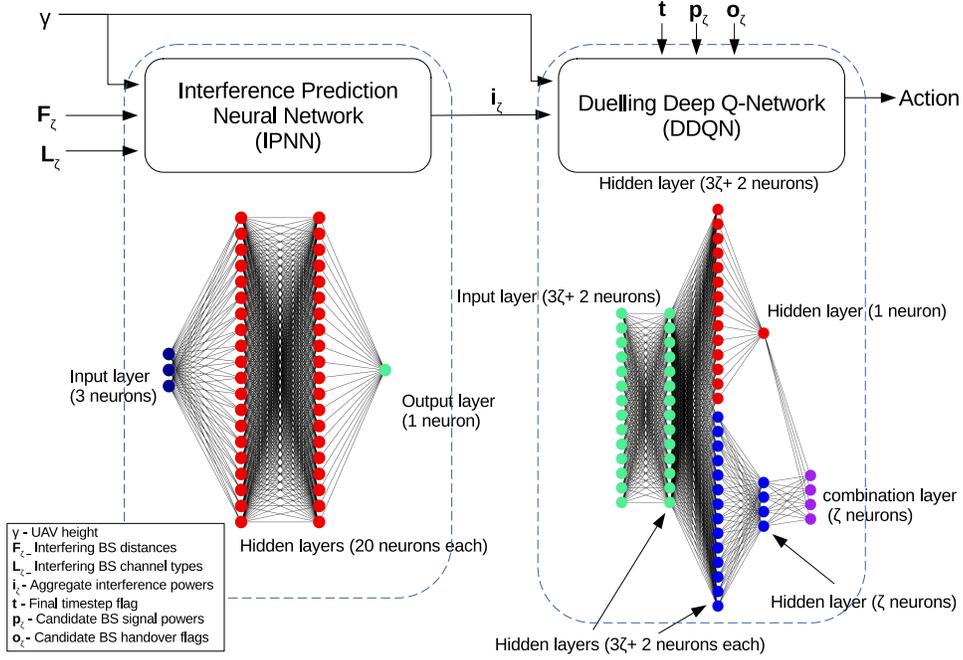}\\
	\vspace{-4mm}
	\caption{
   The structure of our proposed \ac{reqiba} solution, with the number of shown \ac{ddqn} neurons corresponding to a candidate \ac{bs} number $\zeta = 4$.
	\vspace{-9mm}
	}
	\label{fig:diagram}
\end{figure*}



From \Fig{diagram} it is clear that the \ac{reqiba} solution consists of two separate modules, which are detailed below.

\subsubsection{Interference Prediction Neural Network}
Known information about the interference consists of the horizontal distances of the interfering \acp{bs} to the \ac{uav} and their channel types. To estimate the aggregate interference power using this information, we propose a regression neural network circuit which we refer to as the \ac{ipnn}, shown on the left-hand side of \Fig{diagram}. This \ac{ipnn} is trained to estimate the received power from a \ac{bs} given a known horizontal distance to the \ac{uav}, the \ac{uav} height above ground, and whether or not there is an \ac{los} obstruction between the two (from the building topology map available to the \ac{uav}). Providing this trained neural network with the \ac{bs} distances and channel types in the known area $\mathcal{W}$ will allow \ac{reqiba} to estimate the total received interference power when the antenna is aligned with the given candidate \ac{bs}, which it can then pass on to the \ac{ddqn} for candidate \ac{bs} selection.

This \ac{ipnn} circuit consists of an input layer, two dense layers, and an output layer. The dense layers have 20 neurons each, with linear and tanh regularisation functions. The input consists of the matrices $\mathbf{F_\zeta}$ and $\mathbf{L_\zeta}$ and \ac{uav} height $\gamma$. The output consists of a matrix of estimated signal power values for each of the corresponding interfering \acp{bs}. We sum the rows of this matrix to give the final output vector $\mathbf{i}_\zeta= (n_{1,t},n_{2,t},...,n_{\zeta,t})$, where $n_{i,t}$ is the estimated total interference power in the area $\mathcal{W}_i$ that would be experienced by the \ac{uav} if it chooses the $i$-th candidate \ac{bs} at timestep $t$.

\subsubsection{Dueling Deep Q Network}

Having estimated the interference power for each of the candidate \acp{bs}, and knowing their received signal powers and the current \ac{uav} association already, we propose a \ac{ddqn} module to make the decision about which candidate \ac{bs} to associate with for the current timestep. The \ac{ddqn} module is based on model-free \ac{rl}. In \ac{rl}, the \ac{uav} chooses an action based on the observed environmental state at each timestep, based on what it expects will maximise the long-term reward. In a classic \ac{rl} problem a so-called Q-Table is used, which specifies the value of each possible action for a given environment state. As the \ac{uav} takes actions, the \ac{rl} algorithm will observe the resulting action rewards and update the Q-Table accordingly. The Q-Table approach has been shown to be very effective for simple environments and action spaces, but if the environment is very complex then the Q-Table becomes very large and difficult to train. The solution to this is to apply a neural network to approximate the function of the Q-Table, that is, return the estimated Q-values of all possible actions for a given state. We apply a \ac{ddqn} architecture to perform the function of this Q-Table, as shown on the right-hand side of \Fig{diagram}. 


The \ac{ddqn} takes the inputs $\mathbf{p}_\zeta$, $\mathbf{o}_\zeta$, $\gamma$ and $\mathbf{t}$ from the system input and $\mathbf{i}_\zeta$ from the \ac{ipnn}. The Q-value of a state-action pair is the sum of the state value and the action advantage functions; a typical \ac{dqn} estimates the Q-value directly, whereas a \ac{ddqn} contains two parallel streams which estimate the state value and the action advantage functions separately, before combining them together to form the Q-value \cite{Wang2016}. This architecture has been shown to improve the policy evaluation of the neural network compared to the basic \ac{dqn} architecture. The \ac{ddqn} consists of an input layer, followed by a dense layer, followed by a split into two streams. In each stream there are two dense layers. The outputs of the streams are then passed into a combination layer where they are joined together to give $\zeta$ Q-values, one for each possible action (candidate \ac{bs}) for the given state. The output of this \ac{ddqn} is a vector of the Q-values, with the action that has the largest Q-value being selected for the given timestep. Note that we do not explicitly provide any information to the \ac{ddqn} about the handover penalty $\tau$; the negative impact of handovers is to be learned by the \ac{ddqn} over the course of its training, which will enable it to infer the cost of a handover in its decision process.

\section{Training \& Evaluation Methodology}
\label{sec:Methodology}
The environment described in Section \ref{sec:SystemModel} is simulated in the R programming language, using the "Keras" library for the \ac{reqiba} solution \cite{keras}. The environmental parameters of the simulated environment are given in Table \ref{tab:table}, and the \ac{reqiba} hyperparameters are given in Table \ref{tab:table2}. In Section \ref{sec:Results2}, to test the robustness of our approach, we will test the performance of \ac{reqiba} for different environmental parameter ranges; these ranges are shown alongside the baseline values of Table \ref{tab:table}. The hyperparameter values shown in Table \ref{tab:table2} were chosen by us based on prior experimental testing, which suggested their suitability for the \ac{reqiba} scenario.
\begin{table}[t]
\vspace{-3mm}
\begin{center}
\caption{Numerical Result Parameters}
\begin{tabular}{ |c|c|>{\color{black}}c| } 
 \hline
 Parameter & Baseline Value & Test Range \\ 
 \hline
 Carrier Frequency & \unit[2]{GHz} & - \\
 Simulation Area & \unit[5]{km} x \unit[5]{km} & - \\
  Building density $\beta$ & \unit[300]{$/\text{km}^2$} & \unit[100-1000]{$/\text{km}^2$}\\
 Building land coverage $\delta$ & 0.5 & -\\
 Building height scale param. $\kappa$ & \unit[20]{m} & -\\
 \ac{uav} velocity & \unit[10]{m/s} & -\\
  \ac{los} pathloss exponent $\alpha_L$ & 2.1 & -\\
 \ac{nlos} pathloss exponent $\alpha_N$ & 4 & -\\
\ac{bs} transmit power $p$ & \unit[40]{W} & -\\
 Near-field pathloss $c$ & \unit[-38.4]{dB} \cite{Elshaer_2016} & - \\
 \ac{bs} antenna elements $N_t$ & 8 & - \\
 Noise power $\sigma^2$ & \unit[$8\cdot10^{-13}$]{W} \cite{Elshaer_2016} & - \\ 
Handover penalty factor $\tau$ & 0.5 & 0-1 \\
 \ac{bs} height above ground $\gamma_{G}$ &  \unit[30]{m} & -\\
 \ac{bs} density $\lambda$ & \unit[5]{$/\text{km}^2$} & \unit[1-10]{$/\text{km}^2$} \\
 \ac{uav} height $\gamma$ & \unit[100]{m} & \unit[20-200]{m}\\
 \ac{uav} antenna beamwidth $\omega$ & \unit[45]{deg.} & \unit[30-90]{deg.}\\
  MC trials & 2000 & - \\
 Episodes per MC trial & 1 & - \\
 Timesteps per episode $T$ & 100 & - \\
 Timestep duration & 1 second & - \\
 \hline
\end{tabular}
 \label{tab:table}
\end{center}
\end{table}
 
 \begin{table}[t]
\vspace{-3mm}
\begin{center}
\caption{\ac{reqiba} Solution Hyperparameters}
\begin{tabular}{ |c|c| } 
 \hline
 Parameter & Value \\ 
 \hline
 Q-value discount factor & 0.1 \\
 Initial epsilon value $\epsilon$ & 1 \\
 Epsilon decay value & 0.995 \\
 Minimum epsilon value & 0.001\\
 Replay memory size & 10000 entries\\
 Replay batch size & 2048 \\
 Candidate \ac{bs} number $\zeta$ & 10\\
 Interfering \ac{bs} number $\xi$ & 125\\
 \hline
\end{tabular}
 \label{tab:table2}
\end{center}
\end{table}

Before we can evaluate our \ac{reqiba} solution it needs to be trained. As \ac{reqiba} consists of two separate modules it is trained in two stages, which we refer to as offline training and online training. The offline training involves training the \ac{ipnn}. As this module is a regression neural network, it relies on supervised learning, wherein labelled data is presented to the network and it learns the relationship between the input and the output (the label). For our scenario this corresponds to the \ac{ipnn} being presented with a dataset of interfering \ac{bs} distances, channel types, \ac{uav} heights and the resulting received signal powers of those \acp{bs}. In a real-world scenario, this dataset would be generated by having the \ac{uav} fly around an urban environment and measure \ac{bs} signal powers with its directional antenna, while also recording its horizontal distance and channel type. We simulate the generation of this dataset by simulating the urban environment over a number of \ac{mc} trials, with the \ac{uav} positioned at the centerpoint of the environment. In each trial the \ac{uav} points its directional antenna towards a random \ac{bs} and records the signal power observed by the directional antenna, alongside the horizontal distance to the \ac{bs}, its channel type (based on the known building topology) and the height of the \ac{uav}. These measurements populate a dataset which is then used to train the \ac{ipnn}.

Having trained the \ac{ipnn}, we carry out the online training of the \ac{ddqn}, following the procedure shown in the Algorithm 1 pseudocode below. We refer to it as online training, as the \ac{ddqn} is trained during the normal operation of the \ac{uav}, in the typical manner of \ac{rl}. We again simulate a number of \ac{mc} trials with generated urban environments and \ac{uav} travel trajectories. For each \ac{mc} trial the \ac{uav} moves from the start to the end-point in a straight line over $T$ timesteps. We use a technique known as Double \ac{dqn} to train the \ac{ddqn} \cite[Chapter 4]{Francois-Lavet_2018}; this involves creating two \acp{ddqn} during the training process referred to as the "Evaluation \ac{ddqn}" and the "Target \ac{ddqn}". The former is used to evaluate the action values and undergoes training, while the latter is used to select the best action for the next observed state. By separating these operations between the two \acp{ddqn} we allow the algorithm to converge faster during training. At each timestep \ac{reqiba} takes the state inputs, generates the aggregate interference powers via the \ac{ipnn}, then estimates the Q-values via the Evaluation \ac{ddqn}.  We follow an $\epsilon$-greedy training procedure, following which a candidate \ac{bs} is chosen either at random with probability $\epsilon$ or based on the highest Q-value as estimated by the Evaluation \ac{ddqn} with probability $1-\epsilon$. The reward (timestep throughput $\rho_t$) is observed. The state inputs, the action taken, the reward, and the next state inputs are stored in a so-called replay buffer. Once this replay buffer has a sufficient number of entries it is used to train the Evaluation \ac{ddqn}, via uniform sampling of the replay buffer into batches of training data. For each entry in the sampled batch, the algorithm calculates a Q-value for the given state-action pair based on the observed reward, chooses the best action for the next observed state using the Target \ac{ddqn}, and then estimates the highest Q-value of the next state, discounted by a certain amount. These new Q-values are then used to train the Evaluation \ac{ddqn}. The value of $\epsilon$ is decayed by a certain factor at the end of each step, so the training process will randomly explore the environment in the beginning and then rely less and less on random decisions as the \ac{ddqn} becomes more and more trained. At the end of the training episode the Target \ac{ddqn} is updated with the weights of the Evaluation \ac{ddqn}. Once the replay buffer is full, old entries in it are overwritten with new entries.


\newenvironment{algocolor}{%
   \setlength{\parindent}{0pt}
   \itshape
   \color{black}
}{}

\begin{algorithm}
 \begin{algocolor}
\SetAlgoLined
 Init $evalDDQN$, $targetDDQN$, $ReplayBuffer$, $\epsilon$
 
 Init $IPNN$ with offline training weights
 
 \ForAll{MC trials}{
 Generate $Environment$
 
 \While{Step $\leq$ $T$}{
  observe $\gamma$, $\mathbf{F}_{\zeta}$, $\mathbf{L}_{\zeta}$, $\mathbf{t}$, $\mathbf{p}_{\zeta}$, $\mathbf{o}_{\zeta}$, as defined in Fig. 2
  
  $\mathbf{i}_{\zeta}$ = $IPNN(\gamma, \mathbf{F}_{\zeta}, \mathbf{L}_{\zeta})$
  
  State = $[\gamma,\mathbf{i}_{\zeta},\mathbf{t},\mathbf{p}_{\zeta},\mathbf{o}_{\zeta}]$
  
  choose $Action$ randomly with probability $\epsilon$ or based on $evalDDQN(State)$ otherwise
  
  Update $Environment$ with $Association = Action$
  
  observe $Reward = \rho_t$
  
   observe $\gamma$, $\mathbf{F}_{\zeta}$, $\mathbf{L}_{\zeta}$, $\mathbf{t}$, $\mathbf{p}_{\zeta}$, $\mathbf{o}_{\zeta}$
  
  $\mathbf{i}_{\zeta}$ = $IPNN(\gamma, \mathbf{F}_{\zeta}, \mathbf{L}_{\zeta})$
  
  Nextstate = $[\gamma,\mathbf{i}_{\zeta},\mathbf{t},\mathbf{p}_{\zeta},\mathbf{o}_{\zeta}]$

  store $State$, $Action$, $Reward$, $Nextstate$ in $ReplayBuffer$
  
  \If{$size(ReplayBuffer) \geq BatchSize$}{
   randomly sample $Batch$ from $ReplayBuffer$
   
   \ForEach{$Entry \in Batch$}{
   take $State$, $Action$, $Reward$, $Nextstate$ from $Entry$
   
   get $\mathbf{t}$ from $State$
   
   get $NextAction$ from $targetDDQN(Nextstate)$
   
   set $Q(State, Action) = Reward + discount * evalDDQN(NextState,NextAction)*(1-\mathbf{t})$
   }
   train $evalDDQN$ with new Q values
   }
   
   decay $\epsilon$ by $decayFactor$
 }
 
 set $targetDDQN = evalDDQN$
 }
 \caption{DDQN Online Training}
 \end{algocolor}
\end{algorithm}

We propose evaluating our \ac{reqiba} solution in two stages. In the first stage we compare the performance of \ac{reqiba} to the \ac{bs} association solution in our prior work \cite{Galkin_2020}. While this prior solution is designed for a static scenario, it can be applied to a mobile scenario as well. By taking this prior solution as a baseline we quantify the performance gains that \ac{reqiba} can provide. \ac{reqiba} is composed of the \ac{ipnn} and \ac{ddqn} modules which process parts of the state inputs, and both of these modules can be used to make a \ac{bs} association decision in isolation of one another. To verify the performance benefits of the full \ac{reqiba} solution we compare it against the performance of the \ac{ipnn} and \ac{ddqn} modules in isolation. 

In the second evaluation stage we verify the performance of \ac{reqiba} against heuristic \ac{bs} association schemes. As the \ac{uav} has access to important information about the environment, it is capable of making \ac{bs} association decisions by following simple heuristic schemes. In our prior work \cite{Galkin_2020} we demonstrated that the performance improvement from applying \ac{ml} is highly dependent on the environmental conditions, and that under certain circumstances the simple heuristic association schemes may be sufficient for the \ac{uav}. For this reason, we are interested in comparing how \ac{reqiba} performs against heuristics under various environmental conditions. This will give us valuable insight on how the environment can determine the most appropriate type of association policy.

\section{Evaluation Results Against Prior Model}
\label{sec:Results1}

In this section we compare the performance of the two \ac{reqiba} modules, the \ac{ipnn} and the \ac{ddqn}, against the performance of our prior static model from \cite{Galkin_2020}. The purpose of this comparison is two-fold. First, we verify that the \ac{reqiba} solution offers a performance improvement over our prior solution, which does not make use of the \ac{ipnn} and \ac{ddqn} modules. Second, as \ac{reqiba} makes use of two connected modules to make an association decision, we verify that both modules offer measurable performance benefits when working together, to validate our choice of solution. We consider two performance metrics for our comparison: the total throughput over an entire episode, and the handover rate over an episode. This performance comparison is carried out across a range of \ac{uav} heights. For ease of comparison of the episode throughput, we take the total episode throughput of our prior solution as a baseline, and normalise the total episode throughput of the \ac{ipnn} and \ac{ddqn} modules with respect to it. For this comparison we consider three variants of the \ac{ddqn} module: The results labelled "\ac{ipnn}+\ac{ddqn}" use the full \ac{reqiba} solution as described in Section \ref{sec:ML}; the results labelled "\ac{ddqn} (No Int.)" are for the \ac{ddqn} module acting in isolation with no inputs relating to the interference power; and the results "\ac{ddqn} (With Int.)" are for the \ac{ddqn} acting in isolation and taking in the matrices $\mathbf{F_\zeta}$ and $\mathbf{L_\zeta}$ directly. Finally, the results labelled "\ac{ipnn}" show the performance when an association decision is made by choosing the \ac{bs} with the lowest interference power in $\mathbf{i}_\zeta$ as estimated by the \ac{ipnn}, without involving the \ac{ddqn}.

\begin{figure}[t!]
\centering
	\includegraphics[width=.45\textwidth]{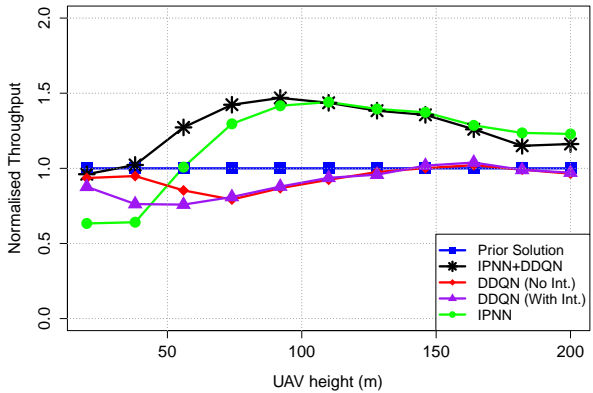}\\
	\vspace{-5mm}
	\caption{
    Normalised throughput showing the performance of the \ac{ipnn} and \ac{ddqn} modules against our prior solution, at different \ac{uav} heights.
	\vspace{-5mm}
	}
	\label{fig:Comparison_throughput}
\end{figure}

\begin{figure}[t!]
\centering
	\includegraphics[width=.45\textwidth]{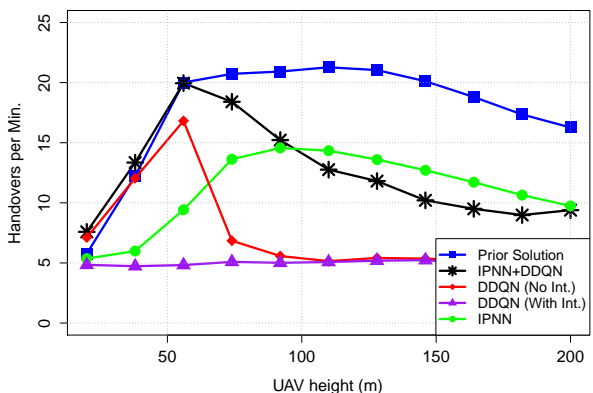}\\
	\vspace{-5mm}
	\caption{
    Handover rate showing the performance of the \ac{ipnn} and \ac{ddqn} modules against our prior solution, at different \ac{uav} heights.
	\vspace{-5mm}
	}
	\label{fig:Comparison_handovers}
\end{figure}

We report that the DDQN-based approaches appear to converge on a solution after approximately 500 MC trials. The plots below show the average performance for the remaining 1500 MC trials. \Fig{Comparison_throughput} shows the resulting throughput, and \Fig{Comparison_handovers} the handover rates. We note that \ac{reqiba} improves the episode throughput by as much as 50\% when compared to the baseline, while offering a significant reduction in the handover rate. This is because \ac{reqiba} offers several improvements over the prior solution. First, the dedicated \ac{ipnn} module is better at estimating the expected interference power than the prior solution, which makes \ac{reqiba} more reliable in its candidate \ac{bs} selection. Second, \ac{reqiba} is able to explicitly learn the negative impact of handovers and take that into consideration by means of its \ac{ddqn} module, whereas the prior solution is designed for a static \ac{uav} scenario, and so ignores the impact of handovers. This causes the prior solution to make an excessive number of handovers during \ac{uav} flight, which negatively impacts the episode throughput. This behaviour suggests that the mobile \ac{uav} connectivity problem cannot be adequately solved by treating it as a sequence of independent static decisions, as the prior solution does.

Comparing the performance of the two modules we see that it is heavily determined by the \ac{uav} height. The \ac{ipnn} in isolation gives very similar throughput improvement as the joint \ac{ipnn}+\ac{ddqn} solution at greater heights, although it performs worse than the baseline at low heights. We explain these observations by the effect of interference at different heights. At low heights interference power is low and the association decision is primarily down to the received signal power from the \acp{bs}, which makes the \ac{ipnn} block not add much value to the decision-making. This results in the \ac{ipnn}+\ac{ddqn} solution performing very similarly to the baseline. As the height increases, the interference starts to play more and more of a role, and so does the \ac{ipnn} module. As a result, the solutions which use the \ac{ipnn} module provide an improvement over the baseline. Note that the baseline solution is capable of inferring some information about interference (albeit not as well as the dedicated \ac{ipnn} module) and so it ends up outperforming the \ac{ddqn} module when the latter is not connected to the \ac{ipnn}. It is interesting to note that passing information about the interfering \acp{bs} directly to the \ac{ddqn} does not improve its performance when compared to not passing it that information; it appears that the \ac{ddqn} is not capable of learning to directly interpret the interference power from the \ac{bs} distances and channel types, and needs the \ac{ipnn} module to perform this function. While the \ac{ddqn} may not be able to provide a good episode throughput without the help of the \ac{ipnn}, it still plays an important role in managing the rate of handovers, as we demonstrate in \Fig{Comparison_handovers}. We observe how the \ac{ipnn}+\ac{ddqn} solution is able to achieve a lower handover rate at greater heights than the pure \ac{ipnn} solution, while still managing a very similar throughput. This demonstrates that while the \ac{ipnn} module by itself may be sufficient for maximising the episode throughput when the \ac{uav} is operating in certain interference-heavy conditions, the \ac{ddqn} module is needed to reduce the resulting rate of handovers. We explore this behaviour further in the next section.

The run-time of the \ac{reqiba} algorithm is tested on a Dell Latitude E5550 laptop, with an Intel i7-8665U CPU at 1.9GHz clock frequency, and 32GB of RAM. The online part of the \ac{ipnn}+\ac{ddqn} solution (both evaluation and training) takes approximately 420ms to run over a single time-step; given that the time-step is modelled as lasting one second we note that the algorithm can feasibly run in real-time on portable hardware.

\section{Evaluation Results Against Heuristic Association}
\label{sec:Results2}
Having verified that our proposed \ac{reqiba} solution provides a significant performance boost over our prior solution, we now evaluate how well \ac{reqiba} compares against heuristic association schemes under different environmental conditions. 
As the problem of \ac{uav} association and handover is new, there is no established performance benchmark to compare against; we therefore perform a comparison against commonly adopted heuristic solutions.
As we observed in the previous section, the \ac{ipnn} module by itself can provide good performance under some circumstances, which is why we include it alongside the full \ac{ipnn}+\ac{ddqn} solution in the following results. The heuristic algorithms selected in this section are detailed below:

\begin{itemize}
    \item Closest \ac{bs} association: As the \ac{uav} has a map of the \acp{bs}, it can determine which \ac{bs} is the closest to it at any given timestep. Under this association scheme the \ac{uav} always connects to the closest \ac{bs}, regardless of the received signal powers or current association. This association is depicted in blue in the figures below.
    \item Highest \ac{sinr} association. The omni-directional antenna on the \ac{uav} can measure the \ac{sinr} of the channel for the strongest signal \acp{bs} around it. While this omni-directional \ac{sinr} will differ from the \ac{sinr} of the directional antenna when aligned, it can still be used to make an association decision directly, instead of using the \ac{reqiba} modules. This type of association corresponds to the usual association policy used by ground users. This association is depicted in red.
    \item Shortest mean distance association. We have assumed in our scenario that the \ac{uav} does not know its trajectory and where it will be in future timesteps. For the sake of a heuristic comparison, we relax this assumption. If the \ac{uav} knows its trajectory over the whole episode, it can choose to connect to the \ac{bs} which has the shortest average distance to the \ac{uav} across all timesteps. This association is depicted in orange.
    \item Angle alignment association. If the \ac{uav} knows the locations of the \acp{bs} and it knows its own trajectory, it is aware of the \ac{bs} whose direction is the closest to the direction that the \ac{uav} is travelling in. To represent a scenario where realigning the directional antenna may be undesirable, we consider an association scheme where the \ac{uav} associates with the \ac{bs} which is the best-aligned with the direction of the \ac{uav} flight. This association is depicted in purple.
\end{itemize}

Unless stated otherwise the results in the figures below are based on the parameter values in Tables I and II. In the following subsections we vary the \ac{uav} height, \ac{bs} density, building density, \ac{uav} antenna beamwidth, and handover penalty, and report on the comparative performance of \ac{reqiba} against the other association schemes. As in the previous section, we normalise the episode throughput of the different association schemes with respect to a baseline, which in this section corresponds to the episode throughput achieved from the closest association scheme. As before, the training appears to converge on a solution after approximately 500 MC trials, with the average performance over the remaining 1500 MC trials shown in the figures. 

\subsection{UAV Height}
\begin{figure}[b!]
\centering
	\includegraphics[width=.45\textwidth]{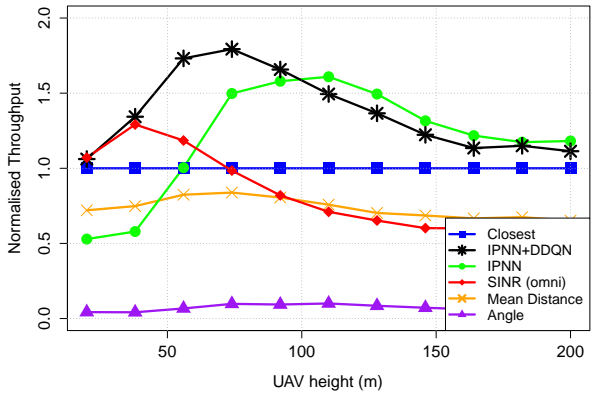}\\
	\vspace{-5mm}
	\caption{
    Normalised throughput showing the performance of our \ac{reqiba} modules, as well as the heuristics, for different \ac{uav} heights.
	\vspace{-5mm}
	}
	\label{fig:Height_throughput}
\end{figure}

In \Fig{Height_throughput} we show the normalised throughput achieved for the different association schemes under varying \ac{uav} heights. We note that the \ac{ipnn}+\ac{ddqn} association scheme outperforms all of the heuristics across the entire range of heights, giving as much as a 70\% throughput improvement over the best heuristic scheme.  As in \Fig{Comparison_throughput}, the \ac{ipnn} association scheme gives poor performance at low heights where interference power is low, and gives good performance at large heights, slightly improving on the \ac{ipnn}+\ac{ddqn} scheme. At low heights the \ac{bs} antenna sidelobe gain plays an important role in the signal performance. As a result, the \ac{ipnn}+\ac{ddqn} association scheme, which takes into account several factors such as interference, candidate \ac{bs} signal strengths, and the cost of handovers, is able to outperform any other association scheme which only considers one factor, while the \ac{ipnn} association scheme performs worse than the simple \ac{sinr} heuristic, despite making use of a trained neural network. At large heights, however, the dominating factor is interference, and choosing a \ac{bs} exclusively based on the resulting interference gives the best throughput.

\begin{figure}[t!]
\centering
	\includegraphics[width=.45\textwidth]{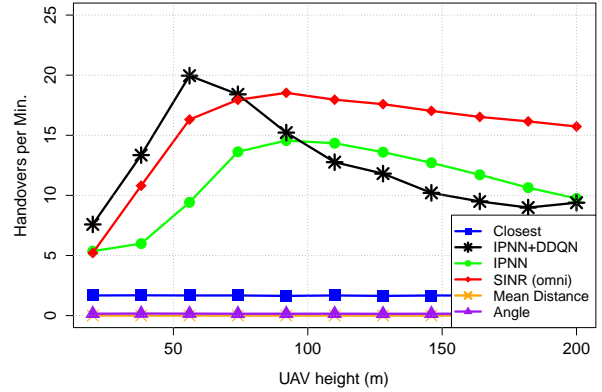}\\
	\vspace{-5mm}
	\caption{
    Handover rate showing the performance of our \ac{reqiba} modules, as well as the heuristics, for different \ac{uav} heights.
	\vspace{-5mm}
	}
	\label{fig:Height_handovers}
\end{figure}

When we consider handover in \Fig{Height_handovers},
we can see that the \ac{ipnn}+\ac{ddqn} gives the largest handover rates at the lower heights, before dropping to values below the \ac{sinr} and \ac{ipnn} association schemes. At low heights, the radio environment is highly volatile, with buildings frequently blocking the \ac{los} channel and strong \ac{bs} sidelobe gains being present. As a result, the \ac{ipnn}+\ac{ddqn} scheme frequently changes the associated \ac{bs}. Because the \ac{sinr} of the steerable directional antenna changes so rapidly, it has a higher handover rate than the \ac{sinr} or the \ac{ipnn} schemes which respond to changes in the omni-directional \ac{sinr} and the interference powers, respectively. As the height increases, the \ac{uav} establishes uninterrupted \ac{los} channels with more \acp{bs} more frequently, which combined with the weaker sidelobe antenna gains result in less channel fluctuations, and therefore less handovers. As the \ac{ipnn}+\ac{ddqn} learns to reduce unnecessary handovers, it results in a lower handover rate than the \ac{sinr} or \ac{ipnn} schemes. On the other hand, the fact that \ac{reqiba}-based association schemes are very dynamic in responding to the changing radio environment means that they still result in a much higher handover rate than the closest \ac{bs} association scheme across the entire range of heights. Considering that at very large heights the \ac{reqiba}-based schemes offer a relatively modest throughput improvement (approximately 20\%) over the closest \ac{bs} association, this significant increase in handover rates may not be justified, in which case it may be worthwhile for the \ac{uav} to rely on simple closest \ac{bs} association when operating at large heights. 

It is also worth noting that the mean distance-based association and angle-aligned association give relatively poor throughput performance, despite benefiting from \textit{a priori} knowledge of the \ac{uav} travel path, which the other association schemes are assumed not to know. The advantage of these association schemes is that they allow the \ac{uav} to pick a single \ac{bs} to connect to and maintain that connection for the entire episode, and so these associations may be useful where limiting the number of handovers is more important than obtaining high throughput.

\subsection{BS Density}
\begin{figure}[t!]
\centering
	\includegraphics[width=.45\textwidth]{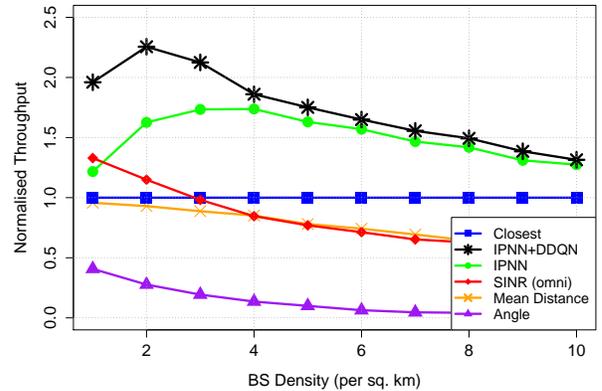}\\
	\vspace{-5mm}
	\caption{
    Normalised Throughput showing the performance of our \ac{reqiba} modules, as well as the heuristics, for different \ac{bs} densities.
	\vspace{-5mm}
	}
	\label{fig:Density_Throughput}
\end{figure}

\begin{figure}[t!]
\centering
	\includegraphics[width=.45\textwidth]{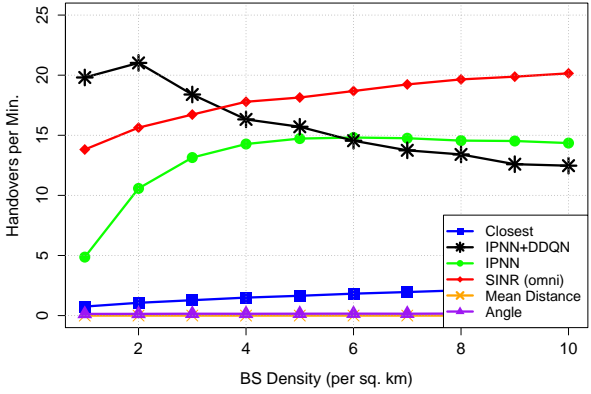}\\
	\vspace{-5mm}
	\caption{
    Handover rate showing the performance of our \ac{reqiba} modules, as well as the heuristics, for different \ac{bs} densities.
	\vspace{-5mm}
	}
	\label{fig:Density_handovers}
\end{figure}

\Fig{Density_Throughput} shows the impact of the \ac{bs} density $\lambda$ on the normalised throughput. At lower densities the \ac{ipnn}+\ac{ddqn} association scheme gives significant improvements over all of the other association schemes, although as the density increases the performance appears to converge to that of the \ac{ipnn} association scheme. This reinforces our observations in the previous sub-section: at low densities a number of factors determine which of the candidate \acp{bs} the \ac{uav} should connect to, whereas as the density increases the interference power becomes the primary deciding factor, which renders the \ac{ipnn}+\ac{ddqn} association marginally better than the \ac{ipnn} association, in terms of throughput. As before, the handover rates in \Fig{Density_handovers} show that the \ac{ipnn}+\ac{ddqn} solution improves  throughput at the expense of a higher handover rate, and that increasing the amount of interference in the environment will result in the \ac{ipnn}+\ac{ddqn} providing a reduced handover rate compared to the pure \ac{ipnn}-based association scheme. Note that the increase in the \ac{bs} density does not impact the run-time of our algorithm, as the architecture of the \ac{ipnn} and \ac{ddqn} modules and the number of inputs are determined by parameters such as $\zeta$ and $\xi$, which are kept constant in our analysis.

\subsection{Building Density}
\begin{figure}[t!]
\centering
	\includegraphics[width=.45\textwidth]{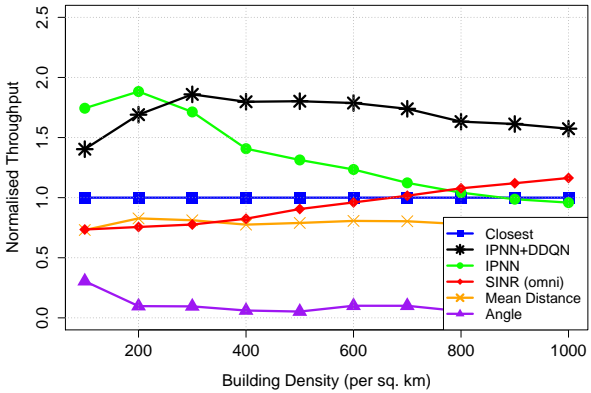}\\
	\vspace{-5mm}
	\caption{
    Normalised Throughput showing the performance of our \ac{reqiba} modules, as well as the heuristics, for different building densities.
	\vspace{-5mm}
	}
	\label{fig:Building_Throughput}
\end{figure}

\begin{figure}[t!]
\centering
	\includegraphics[width=.45\textwidth]{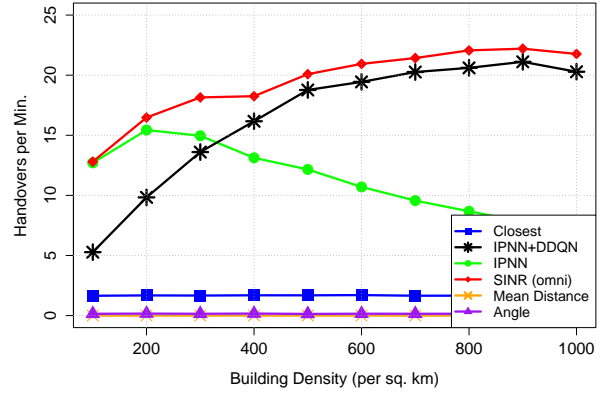}\\
	\vspace{-5mm}
	\caption{
    Handover rate showing the performance of our \ac{reqiba} modules, as well as the heuristics, for different building densities.
	\vspace{-5mm}
	}
	\label{fig:Building_handovers}
\end{figure}

\Fig{Building_Throughput} and \Fig{Building_handovers} show the performance under different densities of buildings in the urban environment. Increasing the building density leads to more \ac{los} blocking obstacles in the environment, which results in wireless channels that fluctuate significantly more as the \ac{uav} moves. The \ac{ipnn} module, as it only considers aggregate interference power, struggles to adapt to this dynamism and so the normalised throughput degrades with increasing density. The \ac{ipnn}+\ac{ddqn} solution is aware of both interference powers as well as candidate \ac{bs} powers, so it is capable of adapting to this increasing channel complexity, and manages to maintain a relatively stable performance improvement over the baseline. As a consequence of reacting to the increasingly dynamic radio environment the \ac{ipnn}+\ac{ddqn} solution sees an increase in the handover rate as the building density increases. We note that very densely built-up environments require very frequent handovers to respond to the volatile radio conditions.

\subsection{UAV Beamwidth}
\begin{figure}[t!]
\centering
	\includegraphics[width=.45\textwidth]{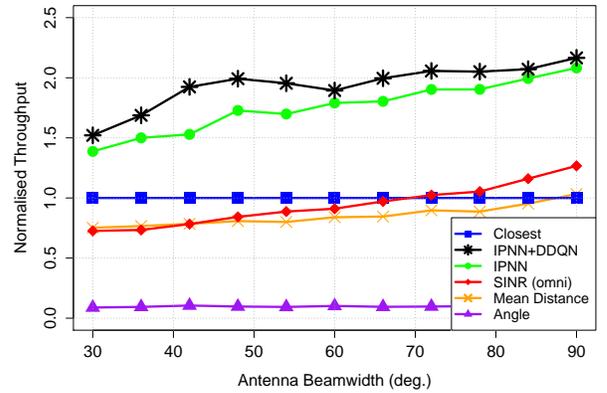}\\
	\vspace{-5mm}
	\caption{
    Normalised Throughput showing the performance of our \ac{reqiba} modules, as well as the heuristics, for different \ac{uav} antenna beamwidths.
	\vspace{-5mm}
	}
	\label{fig:Beamwidth_Throughput}
\end{figure}

\begin{figure}[t!]
\centering
	\includegraphics[width=.45\textwidth]{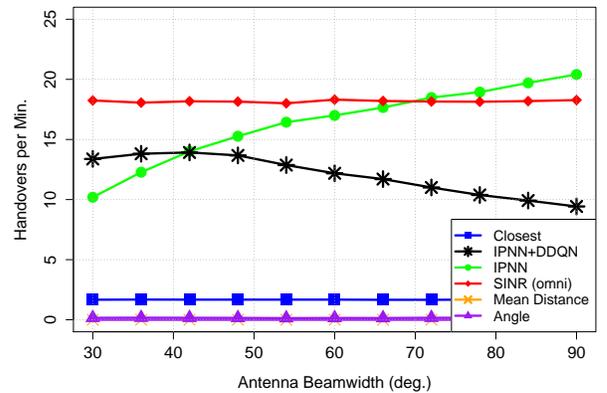}\\
	\vspace{-5mm}
	\caption{
    Handover rate showing the performance of our \ac{reqiba} modules, as well as the heuristics, for different \ac{uav} antenna beamwidths.
	\vspace{-5mm}
	}
	\label{fig:Beamwidth_handovers}
\end{figure}

\Fig{Beamwidth_Throughput} and \Fig{Beamwidth_handovers} show the effects of \ac{uav} antenna beamwidth on the performance.  Increasing the beamwidth of the antenna allows more interfering \acp{bs} to be illuminated by the directional antenna, which  increases the overall interference power. This appears to increase the fluctuations in interference power, as \Fig{Beamwidth_handovers} shows a significant increase in the handover rate of the \ac{ipnn}-based association scheme. By contrast, the \ac{ddqn} association scheme is able to recognise the negative impact of these interference fluctuations and is able to intelligently avoid unnecessary handovers, thus reducing the handover rate as the beamwidth increases. It is interesting to note that the resulting normalised throughput appears to be quite similar for both \ac{reqiba}-based association schemes, as the \ac{ipnn} scheme focuses on improving the channel quality at all costs, while the \ac{ipnn}+\ac{ddqn} may opt for a worse channel, but benefit from the reduced overheads of frequent handovers.

\subsection{Handover Penalty}
\begin{figure}[t!]
\centering
	\includegraphics[width=.45\textwidth]{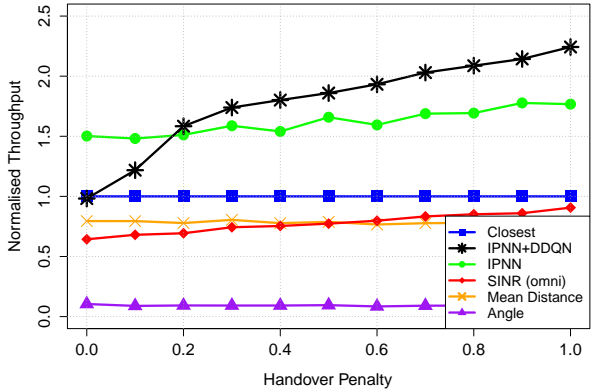}\\
	\vspace{-5mm}
	\caption{
    Normalised Throughput showing the performance of our \ac{reqiba} modules, as well as the heuristics, for different handover penalty factors $\tau$.
	\vspace{-5mm}
	}
	\label{fig:Handover_Throughput}
\end{figure}

\begin{figure}[t!]
\centering
	\includegraphics[width=.45\textwidth]{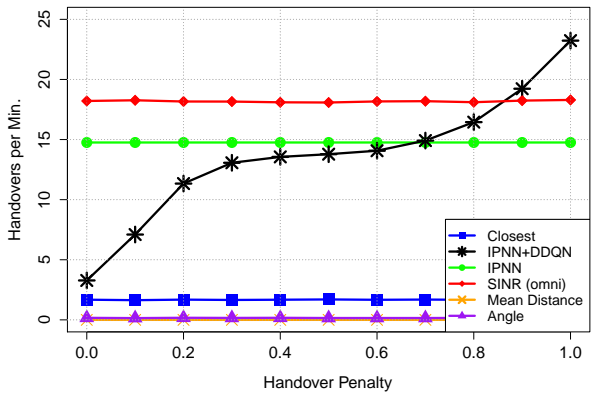}\\
	\vspace{-5mm}
	\caption{
    Handover rate showing the performance of our \ac{reqiba} modules, as well as the heuristics, for different handover penalty factors $\tau$.
	\vspace{-5mm}
	}
	\label{fig:Handover_handovers}
\end{figure}

We now consider the effect of the handover penalty $\tau$ on the \ac{uav} performance. Recall that $\tau$ has a range between 0 and 1, and impacts the received reward in a timestep where a handover occurs. Values of $\tau$ closer to 0 correspond to heavy penalty for carrying out a handover, and this is reflected in the resulting throughput shown in \Fig{Handover_Throughput}. 
The figure shows that the \ac{ipnn}+\ac{ddqn} based association method suffers heavily for low values of $\tau$. The \ac{ddqn} module relies on trial-and-error exploration to learn which actions to take for a given observed state; low values of $\tau$, however, heavily punish any exploration and attempts to connect to better \acp{bs}. This causes the \ac{ddqn} module to learn a very conservative association policy which results in very low throughput performance, much lower than the \ac{ipnn} association scheme, which ignores the impact of handover penalties entirely. It is worth noting that the \ac{ipnn} scheme appears to only suffer minor throughput degradation for lower values of $\tau$, despite the relatively large handover rate, as shown in \Fig{Handover_handovers}. We can see that for low values of $\tau$ the \ac{ipnn}+\ac{ddqn} solution prioritises minimising the handover rate at all costs, while as $\tau$ increases the \ac{ipnn}+\ac{ddqn} association scheme begins to more freely carry out handovers during flight, even exceeding the handover rate of the \ac{ipnn}-based association above a certain point.

\subsection{Randomised Trajectory}
\begin{figure}[t!]
\centering
	\includegraphics[width=.45\textwidth]{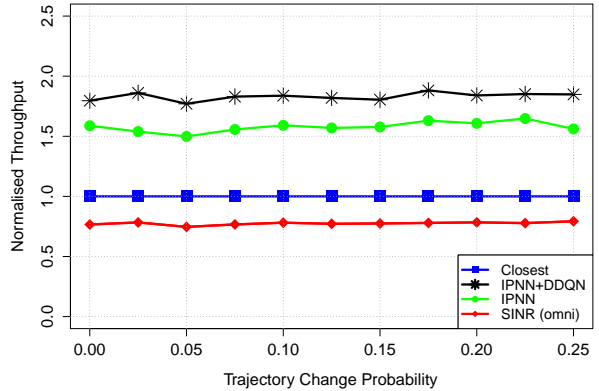}\\
	\vspace{-5mm}
	\caption{
    Normalised Throughput showing the performance of our \ac{reqiba} modules, as well as the heuristics, for different probabilities of a timestep trajectory change.
	\vspace{-5mm}
	}
	\label{fig:Random_Trajectory}
\end{figure}

Up to now, we have assumed that the \ac{uav} flies in a straight line in each episode, as our main focus is to evaluate \ac{uav} base station association rather than path planning. To verify that our \ac{reqiba} solution works for non-linear trajectories as well, we now relax this assumption and consider randomised trajectories in each episode. At each timestep, with a certain probability the \ac{uav} will adjust its horizontal travel vector by 30 degrees either clockwise or counter-clockwise. \Fig{Random_Trajectory} shows the resulting performance as the \ac{uav} trajectory becomes increasingly random in each episode. We note that while the randomness appears to introduce additional noise to the measurements, none of the association algorithms appear to deteriorate with the non-linear \ac{uav} movement. While the Q-Learning component of the \ac{ddqn} does make predictions about future timesteps, as we use low values of the Q-value discount factor the additional randomness does not appear to affect the \ac{ddqn} performance. This indicates that \ac{reqiba} is applicable to a wide variety of \ac{uav} mobility scenarios.

\section{Discussion \& Conclusion}
\label{sec:Conclusion}
In this paper we have proposed an \ac{ml}-based \ac{bs} association scheme referred to as \ac{reqiba} that would allow a \ac{uav} moving through an urban area to intelligently choose which \acp{bs} to connect to, to maximise the overall data throughput during the flight while keeping the rate of handovers manageable. Our proposed solution consists of two modules: a regression neural network module for estimating the aggregate interference powers for each candidate \ac{bs}'s channel, and a \ac{ddqn} module for choosing the candidate \ac{bs} in each timestep, based on \ac{rl} training. Our numerical results show that the \ac{reqiba} solution allows a \ac{uav} to significantly improve its total throughput, when compared to the \ac{ml} solution we proposed for a static \ac{uav} scenario in our prior work \cite{Galkin_2020}, as well as
to heuristic association schemes that use the available environment information. It has been established by the wireless community that \ac{uav} wireless channels are interference-limited due to the low levels of signal blockage; we have shown that under certain conditions the interference power is such an issue that we can achieve the best throughput by simply choosing the candidate \ac{bs} with the lowest interference power on its channel, without taking into account other factors such as the signal power of the candidate \acp{bs} or the effects of handovers. The use of our full \ac{reqiba} solution with the \ac{ddqn} module becomes justified when the environment is less influenced by interference (such as due to lower \ac{bs} density or lower \ac{uav} height above ground), and where achieving the best throughput means balancing a number of environmental factors in the decision process. Even in scenarios where the \ac{ipnn} module alone is sufficient to maximise the throughput, the \ac{ddqn} module plays an important role in managing the rate of handovers, as it explicitly factors in the impact of carrying out a handover to a new \ac{bs}. Without this function, the \ac{uav} may carry out very frequent handovers to respond to the dynamic environmental conditions, in the order of one handover every three or four seconds according to our results.

Our analysis shows that while the \ac{reqiba} solution using the joint \ac{ipnn}+\ac{ddqn} association can offer significant benefits to the \ac{uav}, there are certain important caveats that need to be taken into account by the \ac{uav} operators before choosing it for the \ac{bs} association task. First of all, we have demonstrated that interference power prediction is a mandatory phase of the association decision process; while the \ac{ipnn} module could be used in isolation to make association decisions, the \ac{ddqn} module relies on information about the interference power, and cannot provide good performance without this input. The \ac{ddqn} module was also shown to react negatively to strong handover penalties, as the penalty punishes any exploration carried out by the \ac{ddqn}, which causes the training process to learn to pursue a handover-minimisation scheme, giving relatively poor results. By comparison, the simpler \ac{ipnn}-only association scheme was shown to be more resilient to strong handover penalties, and would be a more appropriate scheme to use in situations where the handover penalties are severe. 

Ultimately, the problem of mobility management involves finding a balance between maximising the channel quality of a moving device while minimising the cost incurred by handovers. Allowing a device the flexibility of choosing its associated \ac{bs} with the changing environment carries the cost of more frequent handovers. If the overheads associated with the handovers are too great, or if the \ac{uav} use-case requires low handover rates, then an \ac{ml}-based association scheme may not be the most appropriate choice in some circumstances. Our results have shown that while certain heuristic association schemes (such as the highest-\ac{sinr} association scheme) are wholly inappropriate for \ac{uav} connectivity, other schemes (such as closest \ac{bs} association) can offer very low handover rates, and therefore may be the most suitable association schemes to adopt in some scenarios.

It should be noted that although our algorithm design was motivated by the unique challenges that \acp{uav} experience when interacting with the cellular network, it can also be applied to terrestrial devices with only minor modifications. These modifications may include the removal of the height above ground input parameter, and the inclusion of a predicted travel vector input (as ground devices, in particular road users, have more predictable trajectories than \acp{uav}, which could be leveraged by our algorithm for its decision-making).

In this work we have considered cellular interference mitigation by means of a steerable, directional \ac{uav} antenna, as per the \ac{3gpp} recommendations. Other interference mitigation techniques are being proposed in the state-of-the-art, such as interference coordination and cooperative beamforming \cite{Mei_2020}. In our future works we may consider a multi-agent optimisation scenario where one agent on the \ac{uav} optimises the association decisions, while another agent applies interference coordination on the side of the \acp{bs}. In future works we may also consider additional \ac{uav} parameters such as flight velocities and trajectories; this would allow us to design intelligent agents which leverage that information in their decision-making. In this work we kept the \ac{bs} heights constant, but the \ac{ipnn} architecture can easily be extended to incorporate variable \ac{bs} heights above ground. We may explore scenarios with variable \ac{bs} heights in future works. Additionally, in this work we have explored a scenario where a single \ac{uav} applied the \ac{reqiba} algorithm on its local hardware to make connectivity decisions; this scenario can be extended to consider a fleet of \acp{uav} on a joint mission. In such a scenario the algorithm may be modified into a centralised algorithm that offloads the connectivity decisions from the individual \acp{uav}. This would reduce the hardware requirements for the \acp{uav}, while introducing latency challenges. Furthermore, the presence of multiple environmentally-aware \acp{uav} may allow for additional data and experience sharing, which could create new AI-driven decision-making problems, and allow for solutions such as federated learning. We may explore these scenarios, and the performance trade-offs introduced by them, in a future work.
\vspace{-1mm}
\section*{Acknowledgement}
\vspace{-1mm}
This material is based upon works supported by the Sci-
ence Foundation Ireland under Grants No. 17/NSFC/5224 and
13/RC/2077\_P2. Prof. DaSilva's participation was supported by the Commonwealth Cyber Initiative (CCI).

\section*{Appendix: CDF of the Instantaneous SINR}
The \ac{cdf} of the instantaneous \ac{sinr} from the serving \ac{bs} at $x_{s,t}$ with channel type $z_{s,t}$ is derived as

\begin{align}
&F_{\sinr}(y) = \Pd\left(\frac{H p \eta(\omega)\mu(\phi_{s,t}) c ((r_{s,t})^2+\Delta \gamma^2)^{-\alpha_{z_{s,t}}/2}}{I+\sigma^2} < y \right) \nonumber\\
&= \Pd\left(H < \frac{y(I +\sigma^2)} {p  \eta(\omega)\mu(\phi_{s,t}) c ((r_{s,t})^2+\Delta \gamma^2)^{-\alpha_{z_{s,t}}/2}} \right) \nonumber \\
& \overset{(a)}{=} \Ed\left[\frac{g\left(m,w(I+\sigma^2)\right)}{\Gamma(m)}\right] \nonumber \\
& \overset{(b)}{=} \Ed\left[1 - \exp(-w(I+\sigma^2))\sum_{k=0}^{m-1}\frac{(w(I+\sigma^2))^{k}}{k!}\right] \nonumber \\
& \overset{(c)}{=} 1 - \sum_{k=0}^{m-1}(-1)^k\frac{w^k}{k!} \Ed\left[\frac{\dr^k \exp(-w(I+\sigma^2)) }{\dr w^k}\right]  \nonumber \\
& \overset{(d)}{=} 1 - \sum_{k=0}^{m-1}(-1)^k\frac{w^k}{k!} \frac{\dr^k \Lc_{(I+\sigma^2)}(w) }{\dr w^k} \nonumber\\
&\overset{(e)}{=} 1 - \sum_{k=0}^{m-1}(-1)^k\frac{w^k}{k!} \sum_{i_{L}+i_{N}+i_{\sigma}=k}\frac{k!}{i_{L}!i_{N}!i_{\sigma}!}\nonumber \\
&\cdot\frac{\dr^{i_{L}} \Lc_{I_{L}}(w)}{\dr w^{i_{L}}}\frac{\dr^{i_{N}}\Lc_{I_{N}}(w)}{\dr w^{i_{N}}}\frac{\dr^{i_{\sigma}}\exp(-w\sigma^2)}{\dr w^{i_{\sigma}}}
\end{align}

\noindent 
where $H$ denotes the Nakagami-m multipath fading component with shape parameter $m$, and $w = y/ p  \eta(\omega)\mu(\phi_{s,t}) c ((r_{s,t})^2+\Delta \gamma^2)^{-\alpha_{z_{s,t}}/2}$. $(a)$ comes from the mulipath fading component having a gamma distribution, with $\Gamma(.)$ and $g(.,.)$ denoting the gamma function and the lower incomplete gamma function, respectively. $(b)$ comes from representing the lower incomplete gamma function as in \cite{Ryzhik_2007}[8.352.2]. $(c)$ comes from the substitution $\exp(-w(I+\sigma^2))(I+\sigma^2)^k = (-1)^k \dr^k \exp(-w(I+\sigma^2))/\dr w^k$. $(d)$ comes from the Leibniz integral rule, where $\Lc_{(I+\sigma^2)}(w)$ is the Laplace transform of $(I+\sigma^2)$. Finally, $(e)$ comes from the fact that $\Lc_{(I+\sigma^2)}(w) = \Lc_{I_L}(w)\Lc_{I_N}(w)\exp(-w\sigma^2)$ following Proposition 3 in \cite{Galkin_2019}, where $I_L$ and $I_N$ are the aggregate interference from \acp{bs} with \ac{los} and \ac{nlos} channels to the \ac{uav}, respectively. It follows then that the k-th derivative of $\Lc_{(I+\sigma^2)}(w)$ can be expressed as a sum of products of higher derivatives of $\Lc_{I_L}(w)$, $\Lc_{I_N}(w)$, and $\exp(-w\sigma^2)$ following the general Leibniz rule.

The aggregate \ac{los} and \ac{nlos} interferences $I_L$ and $I_N$ are random variables which are the sum of the received interference signals from sets of \acp{bs} that have the corresponding channel type to the \ac{uav}. We denote the sets of these interfering \acp{bs} as $\Phi_{\mathcal{W}L}, \Phi_{\mathcal{W}N} \subset \Phi_{\mathcal{W}}$; these are inhomogeneous \ac{ppp} with intensity functions $\lambda_L(x) = \Pd_{los}(x)\lambda$ and $\lambda_N(x)=(1-\Pd_{los}(x))\lambda$, where $\Pd_{los}(x)$ denotes a probability function which gives the \ac{los} probability of a point at $x$ to the \ac{uav} \cite{Galkin_2019}. A number of \ac{los} probability functions for \ac{uav} communications in cities have been proposed by the wireless community including Eq. (2) in \cite{Galkin_2019}, any of these may be applied for the derivation below.

The Laplace transform of $\Lc_{I_j}(w)$ where $j\in\{L,N\}$ is derived as:

\begin{align}
 &\Lc_{I_j}(w) = \Ed\Big[\exp(-wI_j)\Big], j \in \{L,N\} \nonumber \\
 &\overset{(a)}{=} \Ed_{\Phi_{\mathcal{W}j}}\Big[\prod_{x\in\Phi_{\mathcal{W}j}} \Ed_{H}\left[\exp(-H f(x,y_t,w,\alpha_j)\right]\Big] \nonumber \\
 &\overset{(b)}{=}\Ed_{\Phi_{\mathcal{W}j}}\Big[\prod_{x\in\Phi_{\mathcal{W}j}}\Big(\frac{m}{f(x,y_t,w,\alpha_j)+m}\Big)^m\Big]\nonumber \\
 &\overset{(c)}{=}\exp\Bigg(-\int_{\mathcal{W}}\left(1-\frac{m}{f(x,y_t,w,\alpha_j)+m}\right)\lambda_j(x)\dr x \Bigg)
\end{align}

where $(a)$ stems from the aggregate interference $I_j$ consisting of a sum of the received signal powers from the \acp{bs} belonging to the set $\Phi_{\mathcal{W}j}$, with each signal affected by Nakagami-m multipath fading, where $f(x,y_t,w,\alpha_j)=w  p  \eta(\omega)\mu( \arctan(\Delta \gamma/||x-y_t||)) c (||x-y_t||^2+\Delta \gamma^2)^{-\alpha_{j}/2}$. $(b)$ comes from the power-series representation of the Nakagami-m multipath fading component. $(c)$ stems from the probability generating functional of a \ac{ppp} \cite{Haenggi_2013}. Note that the integral is over the area of $\mathcal{W}$, as all interfering \acp{bs} outside of the illuminated area $\mathcal{W}$ will give 0 interference due to the \ac{uav} directional antenna gain.
For more information on the mathematical characterisation of \ac{uav}-\ac{bs} connectivity, the reader is referred to our prior works \cite{Galkin_2018,amer2020performance,Galkin_2019}.
\ifCLASSOPTIONcaptionsoff
  \newpage
\fi



\bibliographystyle{./IEEEtran}
\bibliography{./IEEEabrv,./IEEEfull}

\vspace{-13mm}
\begin{IEEEbiography}[{\includegraphics[width=1in,height=1.25in,clip]{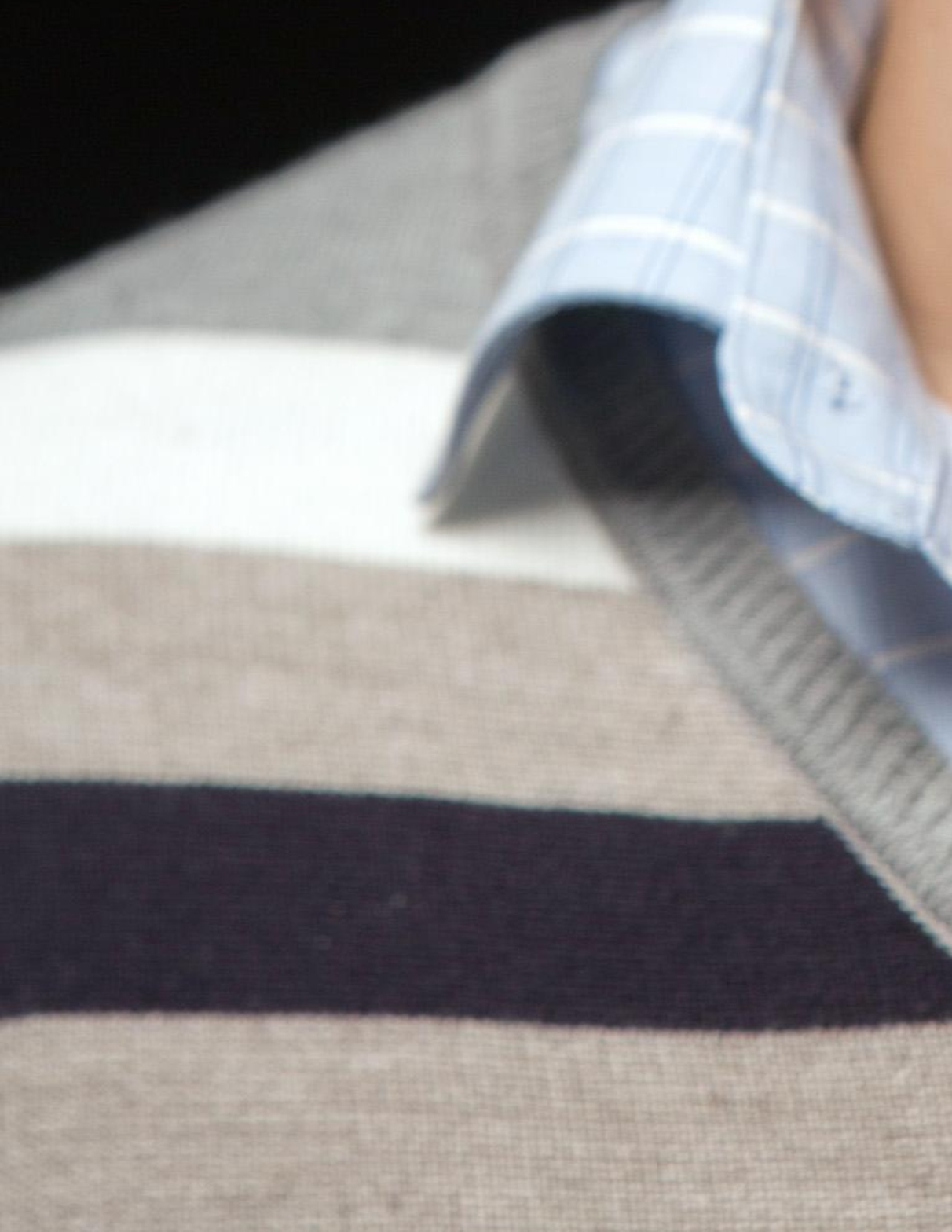}}]{Boris Galkin}
is currently a research fellow in CONNECT, Trinity College Dublin, Ireland. He was awarded a BAI and MAI in computer \& electronic engineering in 2014 and a Ph.D. degree in 2019 from Trinity College Dublin, Ireland. His research interests include unmanned aerial vehicles, heterogeneous networks, and the application of AI and Machine Learning to 5G and Beyond networks.
\end{IEEEbiography}
\vspace{-12mm}
\begin{IEEEbiography}[{\includegraphics[width=1in,height=1.25in,clip]{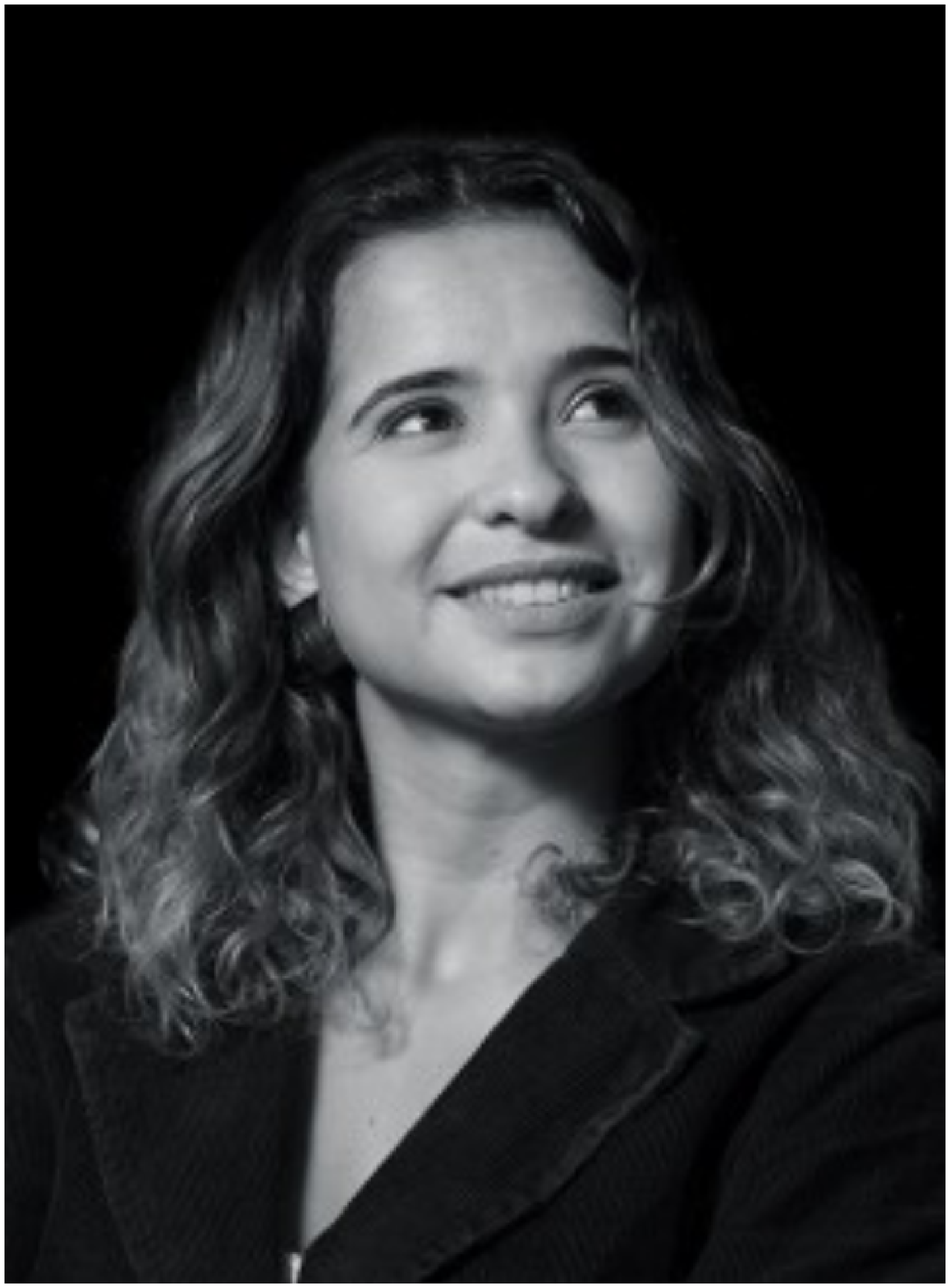}}]{Erika Fonseca}
is a PhD researcher at CONNECT in Trinity College Dublin. She received her MSc degree in Network Computing from University Federal Fluminense in 2017 and the BSc degree in Telecommunications Engineering in 2013, also from the University Federal Fluminense. She has experience in wireless network research and her research interest is focused on 5G cellular networks and software-defined radio.
\end{IEEEbiography}

\vspace{-12mm}
\begin{IEEEbiography}[{\includegraphics[width=1in,height=1.25in,clip]{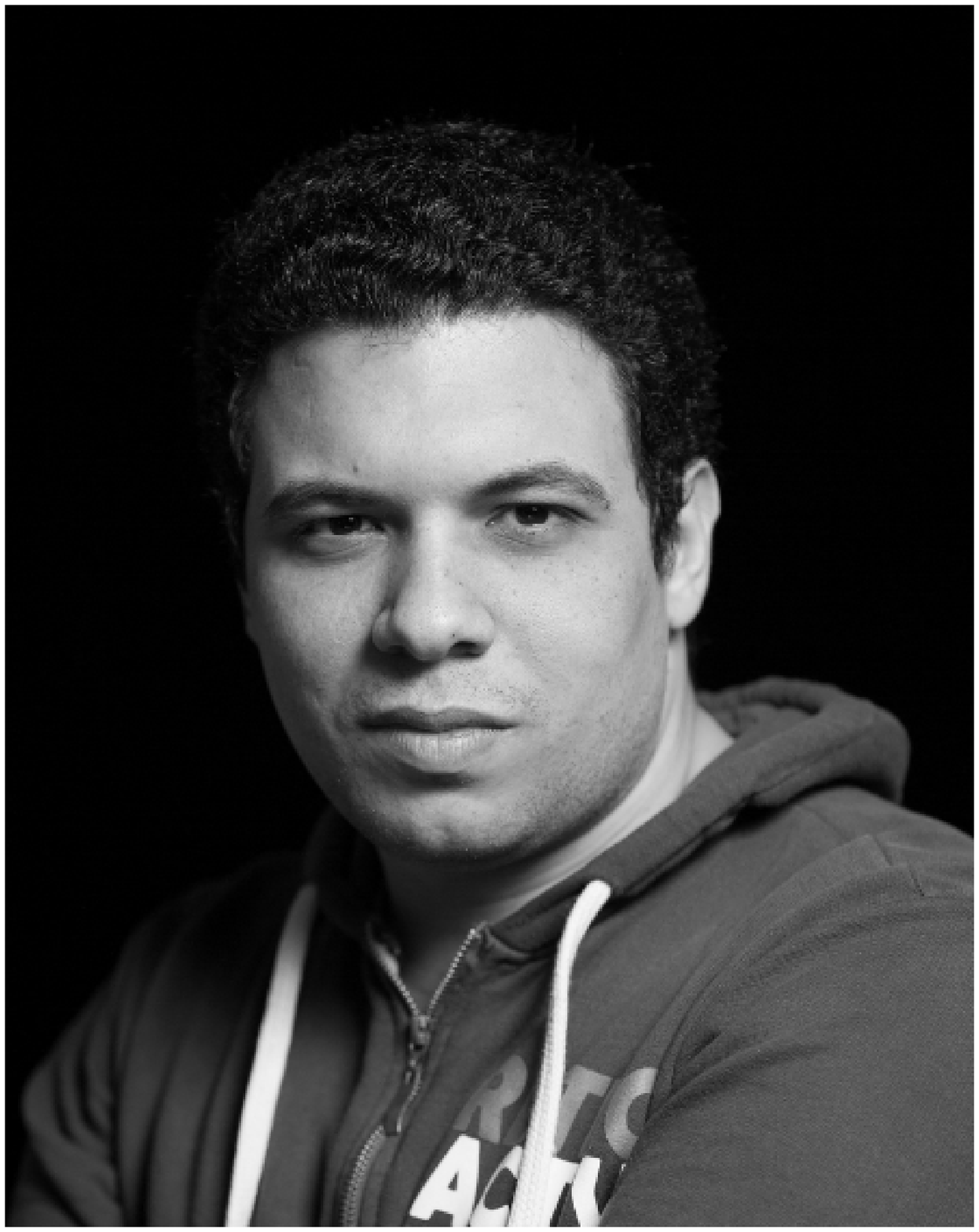}}]{Ramy Amer}
is a PhD researcher at CONNECT in Trinity College Dublin. He received the BSc and MSc degrees in Electrical Engineering from the Alexandria University, Egypt, in 2010, and 2016, respectively. His research interests include cross layer design, cognitive radio, wireless caching, and energy harvesting. 
\end{IEEEbiography}

\vspace{-12mm}
\begin{IEEEbiography}[{\includegraphics[width=1in,height=1.25in,clip]{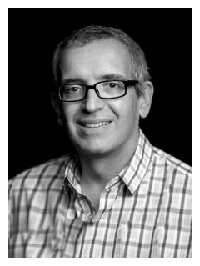}}]{Luiz A. DaSilva}
is the Executive Director of the Commonwealth Cyber Initiative (CCI), and Bradley Professor of Cybersecurity at Virginia Tech. He was previously the Professor of Telecommunications (personal chair) at Trinity College Dublin. Until March 2020, he served as the Director of CONNECT, the telecommunications research centre funded by the Science Foundation Ireland. His research focuses on distributed and adaptive resource management in wireless networks, and in particular cognitive radio networks and the application of game theory to wireless networks.
\end{IEEEbiography}
\vspace{-12mm}
\begin{IEEEbiography}[{\includegraphics[width=1in,height=1.25in,clip]{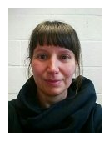}}]{Ivana Dusparic}
is an Ussher Assistant Professor in the School of Computer Science and Statistics at Trinity College Dublin since 2016. She holds a BSc from La Roche College, PA, USA (2003), and MSc (2005) and PhD (2010) from Trinity College Dublin. Her expertise and research interests lie in the use of Artificial Intelligence techniques (reinforcement learning, multi-agent systems) to achieve autonomous optimization of large-scale heterogeneous infrastructures, applied to smart cities and sustainable urban mobility.
\end{IEEEbiography}

\end{document}